\documentclass[10pt,a4paper]{article}
\usepackage{natbib}

\usepackage{amsmath, amssymb, amsfonts}
\usepackage[amsmath,thmmarks,amsthm]{ntheorem}
\usepackage{enumerate}
\usepackage{graphicx}
 
\usepackage{verbatim}
\usepackage{url}

\usepackage{graphicx}
\usepackage{color}

\usepackage{amsmath}
\usepackage{amsfonts}
\usepackage{enumerate}
\usepackage{verbatim}
\usepackage{amssymb}
\usepackage{setspace}
\usepackage{epsfig}
\usepackage{url}
\usepackage{graphicx}

\usepackage{amsmath}

\usepackage[dvipsnames]{xcolor}

 \newcommand{\Int}{\operatorname{int}}
 \newcommand{\Clos}{\operatorname{clos}}

 \newcommand{\diag}{\operatorname{diag}}
 
 \newcommand{\tr}{\operatorname{tr}}

\newcommand*\diff{\mathop{}\!\mathrm{d}}

 \newtheorem{Theorem}{Theorem}
 \newtheorem{Lemma}{Lemma}

 \newtheorem{Corollary}{Corollary}
  \newtheorem{Definition}{Definition}
 
 \newtheorem{Proposition}[Theorem]{Proposition}
 \newtheorem{Remark} {Remark}
 \newtheorem{Example} {Example}

\newcommand {\R}{\mathbb R}

\newcommand {\M}{\mathbb M}

\newcommand{\be}{\begin{equation}}
\newcommand{\ee}{\end{equation}}
\newcommand{\sgn}{\operatorname{{\mathrm sgn}}}
\newcommand{\V}{\mathcal V}

\newcommand{\updt}[1]{{\color{black}#1}}

\newcommand*\dif{\mathop{}\!\mathrm{d}}

\begin{document}

%\begin{frontmatter}
\title{A Generalization of Linear Positive Systems with Applications to Nonlinear Systems: Invariant Sets and the Poincar\'{e}-Bendixson Property\thanks{This research  was partially supported by  research grants from  the Israel Science Foundation    and the
 Binational Science Foundation. An abridged version of this paper has appeared  in Proc.~2019
Mediterranean Conf. on Control and Automation (MED'2019).}}
 \author{Eyal Weiss and Michael Margaliot\thanks{E. Weiss is with the Dept. of Computer Science, Bar-Ilan University, Ramat Gan, 5290002, Israel. 
 	M. Margaliot (Corresponding Author) is with the Department of  Elec. Eng.-Systems and the Sagol School of Neuroscience,  Tel-Aviv
 	University, Tel-Aviv 69978, Israel. E-mail:  \texttt{michaelm@tauex.tau.ac.il}}}
%%%%%%%%%%%%%%%%%%%%%%%%%

\pagestyle{empty}
\maketitle
\thispagestyle{empty}
%\doublespace 
%%%%%%%%%%%%%%%%%%%%%%%%%%%%%%%%%%%%%%%%%%%%%%%%%%%%%%%%%
 %\begin{multicols}{2} 
 
\begin{abstract}
%%%%%%%%%%%%%%%%%
The dynamics of linear positive systems map the   positive orthant to itself. 
In other words, it maps a  set of vectors with zero sign variations to itself. 
This raises the following question: what linear systems map the set of 
vectors with~$k$ sign variations to itself? 
We address  this question using tools from the theory of cooperative dynamical systems
and the theory of totally positive matrices. This yields a generalization of positive 
linear systems called~$k$-positive linear systems, that reduces to positive systems for~$k=1$.
 We describe   applications of this new type of systems
to the analysis of nonlinear   dynamical systems. In particular, we 
show that such systems  admit certain explicit  
 invariant sets, and for the case~$k=2$  establish  the
Poincar\'{e}-Bendixson property for any bounded trajectory. 
 %%%%%%
\end{abstract}
%%%%%%%%%%%%%%%%%%%%%%%%%%%%%%%%%%%%%%%%%%%%%%%%%%%%%%%%%%%%%%%%%%%%%%%%%%%%

%\begin{keyword}%%
%%%%%%%%%%%%%%%%%%%%%%%%%%%%%%%%%%%
%Totally positive matrices, asymptotic stability, Poincar\'e-Bendixson property, 
%sign variation diminishing property,  cyclic feedback systems, 
% compound matrices. 
%%%%
%\end{keyword}
%%%%%%%%%%%%%%%%
%\end{frontmatter}

\section{Introduction}
%%%%%%%%%%%%%%%%%%%%%%%%
% linear positive systems and relation to sign-variations
  Positive dynamical systems 
  arise naturally   in many fields of science where the state-variables represent quantities that can only take nonnegative values~\citep{farina2000}.
  For example, in compartmental systems~\citep{sandberg78} every state-variable represents the density of ``particles'' in a compartment, and this cannot be negative. 
	In chemical reaction networks the state-variables represent 
	 reactant concentrations.
	Another important example
	are models describing the evolution of
	probabilities (e.g. Markov chains)~\citep{gunter2017}.

The dynamics of such systems map the nonnegative orthant 
\[
\R^n_+:=\{x\in\R^n: x_i\geq 0 \text{ for all }i \} 
\]
 to itself 
(and also~$\R^n_-:=-\R^n_+$ to itself).
Intuitively speaking, 
the dynamics  map vectors with zero sign variations to vectors with zero sign variations. 

 In this paper, we suggest a generalization called a~\emph{$k$-positive 
linear system}. Such a system maps the set of vectors with at most~$k-1$ sign variations to itself. 
For the case~$k=1$ this reduces to a positive linear system. 
%%%%%%
But for~$k\geq 2 $ the system may be~$k$-positive even if it is not 
a positive system in the usual sense. 

Positive linear systems are important in their own right, 
and are an active area of research (see, e.g. the recent tutorial by~\cite{posi-tutorial}),
 but   also  play an important role  in the context of nonlinear systems. 
Indeed, if the variational system  associated with the 
 nonlinear system (see the exact definition below) is a positive linear time-varying~(LTV) system
then the nonlinear system is \emph{cooperative} and this has far 
reaching consequences~\citep{hlsmith}. We generalize this by defining~\emph{$k$-cooperative systems} as systems with a  variational system that is a~$k$-positive LTV. 
We describe the implications of this on the asymptotic behavior of the nonlinear system. 
In particular, we  strengthen a seminal result of~\cite{sanchez2009cones}
to prove the Poincar\'{e}-Bendixson property for 
any trajectory  of a~$2$-cooperative system  that remains in a compact set. Note that for our special case we are able to use the nested structure of the invariant sets of a~$2$-cooperative system  to prove a result that is considerably
stronger than the one in~\citep{sanchez2009cones}. 
We believe that these 
 results provide new tools for analyzing the asymptotic behavior
 of nonlinear dynamical systems. For a recent application to an important closed-loop system from systems biology, see~\cite{Margaliot868000}.

We begin with motivating 
 the general ideas in a slightly simplified setting. More general and rigorous statements are given in the next sections.  For a matrix~$B\in\R^{n\times m}$ 
we write~$B\geq 0$ [$B\gg 0$] if every entry of~$B$ is nonnegative [positive]. 
Recall that a matrix~$P\in\R^{n\times n }$
is called \emph{Metzler} if every off-diagonal entry of~$P$ is nonnegative.

Consider the  LTV  system
\be\label{eq:ltv}
\dot x(\tau)=A(\tau)x(\tau),\; x(t_0)=x_0,
\ee
with~$A:(a,b)\to\R^{n\times n}$ a continuous matrix function. 
The associated LTV matrix 
differential system  is:
%%%%%%%%%%%%%%%%%%%%%%%%%%%%%%%
\be\label{eq:matltv}
\dot \Phi(\tau)=A(\tau)\Phi(\tau),\; \Phi(t_0)=I.
\ee
Recall that for any pair~$(t_0,t)$ the solution~$x(t)$
 of~\eqref{eq:ltv} at time~$t$
is given by~$x(t)=\Phi(t,t_0)x(t_0)$, 
where~$\Phi(t,t_0)$ is the solution of~\eqref{eq:matltv} at time~$t$. 
We refer to~$\Phi(t,t_0)$ as the \emph{transition matrix}
 from time~$t_0$ to time~$t$ of~\eqref{eq:ltv}. 

The system~\eqref{eq:ltv} is said to be  positive on the time interval~$(a,b)$ 
if  for any pair~$(t_0,t)$ with~$a<t_0< t<b$   
and any~$x(t_0)\in\R^n_+$ we have~$x(t)\in\R^n_+$.
Equivalently,~$\Phi(t,t_0)\geq 0$ 
   for all~$a<t_0<t<b $. 
It is well-known that
this holds if and only if~(iff) $A(\tau)$ is   Metzler   for
 all~$  a<\tau<b$. 
Thus, we have the following set of equivalent conditions:
\begin{itemize}
%%%%%%%%%%%%
\item The LTV~\eqref{eq:ltv} is positive on the
 time interval~$(a,b)$; 
%%%
\item All the minors of order one of~$\Phi(t,t_0)$ are nonnegative for all~$a<t_0<t<b$; 
%%%%%
\item $A(\tau)$ is Metzler for all~$a<\tau<b$.
\end{itemize}
  
Our goal here is to introduce a generalization called a~\emph{$k$-positive system}.
This is an~LTV that maps the set of vectors with at most~$k-1$ sign variations 
 to itself. In particular, 
the  standard   positive system is a~$1$-positive system.
We show that the following is a 
  set of equivalent conditions:
\begin{itemize}
\item The LTV~\eqref{eq:ltv} is~$k$-positive on the time interval~$(a,b)$; 
%%%
\item All the minors of order~$k$ of the transition matrix~$\Phi(t,t_0)$
 are nonnegative for all~$a<t_0<t<b$; 
%%%%%
\item $A^{[k]}(\tau)$ is Metzler for all~$a<\tau<b$.
\end{itemize}
Here~$A^{[j]}(\tau)$ denotes  the~$j$'th additive compound of~$A(\tau)$ (see e.g.,~\cite{muldo1990}). In particular~$A^{[1]}=A$, so for~$k=1$ we obtain the set of conditions described above for a positive~LTV. 
We provide for every~$k$ a simple
 condition on the structure of~$A(t)$ guaranteeing that~$A^{[k]}(t)$ is Metzler. Thus, 
our results do not require computing the transition matrix. 
Specifically, we show that
an LTV  is~$(n-1)$-positive  iff it is a competitive system
(up to an appropriate coordinate transformation).
For~$1<k<n$  
it is~$k$-positive with~$k$ even 
iff it is~$2$-positive, and it 
is~$k$-positive with~$k$ odd 
iff it is~$1$-positive.

Positive LTVs play an important role in the analysis of time-varying
\emph{nonlinear} dynamical systems. To explain this, consider the   time-varying nonlinear system:
%%%%%%%%%%%%%%%%%%%%%%%%%%%
\be\label{eq:nlin}
\dot x(t)=f(t,x(t)),
\ee
whose trajectories evolve on a convex
state-space~$\Omega\subseteq\R^n$. Assume that~$f$ is~$C^1$ with respect 
 to~$x$, and denote its Jacobian  by~$J(t,x):= \frac{\partial }{\partial x} f(t,x)$.
%%%
For~$p\in\Omega$, let~$x(t,p)$ denote the solution of~\eqref{eq:nlin} at time~$t$ with~$x(0)=p$. For~$p,q\in\Omega$, let
\[
z(t):=x(t,p)-x(t,q),
\]
that is, the difference at time~$t$ between the solutions 
emanating at time zero from~$p$ and from~$q$. Then
\begin{align}\label{eq:var}
\dot z(t) &=A^{pq}(t)z(t),
\end{align}
where~$ A^{pq}(t) :=\int_0^1  J(t,rx(t,p)+(1-r)x(t,q) ) \diff r $.
Eq.~\eqref{eq:var}
  is  called  the \emph{variational system}, as it
describes how a variation in the initial condition evolves
 with time. 
 
If~$A^{pq}(t)$ is Metzler  for all~$t\geq 0$ and all~$p,q\in\Omega $  
then~\eqref{eq:var} is a positive~LTV. Then we conclude that
\be\label{eq:impli} 
p\leq q \implies x(t,p)\leq x(t,q) \text{ for all } t\geq 0,
\ee
i.e.,~\eqref{eq:nlin} is a cooperative dynamical system. 
Note that if~$0\in\Omega$ and~$0$ is an equilibrium point of~\eqref{eq:nlin}
then~\eqref{eq:impli}  implies that~$\R^n_+$ is an invariant set of~\eqref{eq:nlin}.
%%%%%
Cooperative systems have a well-ordered behavior. For example, in the time-invariant case and when the state-space~$\Omega$ is compact almost every trajectory converges to an equilibrium point~\citep{hlsmith}.

Intuitively speaking,~\eqref{eq:impli} can be stated as follows:
if~$p-q$ has zero sign variations then~$x(t,p)-x(t,q)$ has zero sign variations for all~$t\geq 0$. 
We call~\eqref{eq:nlin} a \emph{$k$-cooperative}
 system if the associated variational system is~$k$-positive. 
This means that
if~$p-q$ has no more than~$k-1$ sign variations then
so does~$x(t,p)-x(t,q)$ for all~$t\geq 0$. 
We then describe the implications of this to the solutions of~\eqref{eq:nlin}. 
In particular, we show that such systems admit special 
invariant sets, and   that~$2$-cooperative systems satisfy a Poincar\'{e}-Bendixson property.

The remainder of this paper is organized as follows. 
The next section reviews    definitions and tools from the theory
of totally positive matrices that are needed later on.  
These include in particular the rigorous
definitions of the number of sign
variations in a vector, the variation diminishing properties
 of   sign-regular matrices, and 
  compound matrices. The next four  sections describe our main results. 
%%%%%
Section~\ref{sec:def_k_pos}
 defines the new notions of a~$k$-positive and a strongly~$k$-positive LTV as systems that
leave certain sets invariant. 
Section~\ref{sec:expa} provides explicit conditions for a system to 
be $k$-positive. Section~\ref{sec:geom} analyzes the geometrical structure of the invariant sets
of~$k$-positive systems, and shows that they are solid cones that include a   linear subspace of dimension~$k$,
but no linear subspace of a higher dimension.
However, these cones are not necessarily convex. 
%%%%%%%%%%%%%%%%%%%%%%%%%%%%%%%%%%%%%%%%%%%%%%%%%%%%%%%%%%
Applications
 to nonlinear systems are given in Section~\ref{sec:applnon}.
  We show that if the variational system associated with 
	the nonlinear system is~$k$-positive then the nonlinear system
	admits certain invariant sets that can be described \emph{explicitly}.
  %%%%%%
	Invariant  sets
play 
 a significant role in many control-theoretic and engineering applications
(see e.g., the survey by~\cite{blanchini1999set} 
and the more recent PhD thesis by~\cite{song2015optimization}), yet 
analytic verification that a set is invariant is a non-trivial
 problem~\citep{horvath2016invariance}.  We also show  
 that~$2$-cooperative systems satisfy a Poincar\'{e}-Bendixson property:
a nonempty compact omega limit set 
 which does not contain any equilibrium points is a closed orbit.
The final section concludes and describes topics for further research.

  We use small   letters to denote column vectors, and capital letters
	to denote matrices. 
For a matrix~$A\in\R^{n\times m}$,~$A'$   denotes the transpose of~$A$.
For a vector~$y\in\R^n$,~$y_i$ is the~$i$'th entry of~$y$.
For two integers~$i\leq j $ we use the notation~$[i,j]$ for the set~$\{i,i+1,\dots,j\}$. 
  For a set~$S$,
 $\Int (S)$ is the interior of~$S$, and~$\Clos(S)$ denotes  its closure.
 For a square matrix~$A$, $\tr(A)$ is the trace of~$A$. 
For~$v_1,\dots,v_n\in\R$, we use~$\diag(v_1,\dots_,v_n)$ to denote
 the  diagonal matrix with diagonal entries~$v_1,\dots,v_n$.
 
%%%%%%%%%%%%%%%%%%%%%%%%%%%%%%%%%%%%%%%%%%%%
\section{Preliminaries} 
%%%%%%%%%%%%%%%%%%%%%%%%%%%%%%%%%%%%%%%%%%%%

We begin by reviewing linear mappings that do not increase the number of sign variations in a vector. 
%%%%%%%%%%%%%%%%%%%%%%%%%%%%%%%%%%%%%%%%%%%%
\subsection{Number of sign variations in a    vector}
%%%%%%%%%%%%%%%%%%%%%%%%%%%%%%%%%%%%%%%%%%%%%%%

For a vector~$y\in\R^n$ with 
no zero entries the number of sign variations
in~$y$ is
\[
%%%%%%%%%%%
\sigma(y):=\left|\{i \in \{1,\dots,n-1\} : y_i y_{i+1}<0\} \right | .
\]
For example, $\sigma(\begin{bmatrix}  -4.2 & 3 & -0.5 \end{bmatrix}')=2$.

In the more general case where the vector
 may include zero entries, we recall
 two 
definitions for the number of sign variations from
   the theory of totally positive matrices.
%%%
For~$y\in\R^n$,  $s^{-}(y)=0$ if~$y =0$, and otherwise
  $s^-(y):=\sigma(\bar y)$, where~$\bar y$ is 
the vector obtained from~$y$ by deleting all its zero entries.
Let~$s^+(y):=\max_{z\in S(y)}\sigma(z)$,
where~$S(y)$ includes all the vectors obtained  by 
	replacing every zero entry in~$y$
	 by either~$+1$ or~$-1$. 
	For example, for
	\be\label{eq:vecyy}
	y=\begin{bmatrix} -1& 1 & 0 &0 &  -3.5  \end{bmatrix}',
	\ee
	$s^-(y)=\sigma (\begin{bmatrix} -1& 1  &  -3.5  \end{bmatrix}' )= 2$,
	and~$s^+(y)=\sigma( \begin{bmatrix} -1& 1 & -1 &1 &  -3.5  \end{bmatrix}' )=4$. 
	It follows from these definitions that
	\[
	0\leq s^-(y) \leq s^+(y)\leq n-1 \text{ for all } 
	 y\in\R^n.  
	\]
		% Let~$	\W:=\{y\in\R^n:s^-(y)=s^+(y)\}$, 
%We note that other notations exist in the literature for the number of sign variation in a vector 
%(see e.g., \cite{oliva1993diffeomorphisms}).

Let
\[
 \V:=\{ x\in\R^n:s^-(x)=s^+(x)\}.
\]
It is not difficult to show that 
\begin{align}\label{eq:defvg}
%%%%
\V&=\{x\in\R^n: x_1\not =0,x_n\not =0, \\&\text{ if } x_i=0 \text{ for some } 
i\in[2,n-1] \text{ then } x_{i-1}x_{i+1}<0\}. \nonumber
\end{align}
For example, for~$n=3$ the vector~$x:=\begin{bmatrix}1&\varepsilon&-1 \end{bmatrix}'$
satisfies~$s^-(x)=s^+(x)$ for all~$\varepsilon\in\R$,
and~$x$ satisfies   condition~\eqref{eq:defvg} for all~$\varepsilon\in\R$.

There is a useful duality relation between~$s^-$ and~$s^+$. 
Let~$D:=\diag (1,-1,\dots,(-1)^{n-1})$. Then (see e.g.~\cite[Ch.~3]{pinkus})
\begin{equation} \label{eq:s_minus_s_plus}
s^{-}(x)+s^{+}(Dx)=n-1 \text{ for all }   x \in \R ^n.
\end{equation}
%%%%
For example, for~$n=5$ and the vector~$y$ in~\eqref{eq:vecyy}, 
we have~$s^-(y)=2$, $s^+(Dy)=s^+(\begin{bmatrix} -1& -1 & 0 &0 &  -3.5  \end{bmatrix}' )=2$, so~$ s^{-}(y)+s^{+}(Dy)=4$.

Next we  review matrices~$A$ satisfying that~$Ax$
has no more sign variations than~$x$. 
%%%%
%%%%%
\subsection{Sign regularity and the variation diminishing property}\label{subsec:SR}
%  SR properties and VDP
Consider a matrix~$A\in\R^{n\times m}$, and pick~$k\in [ 1,\min(n,m)]$.
The matrix is said
 to be
   \emph{sign-regular of order~$k$} (denoted~$SR_k$) if all its minors of order~$k$ are nonnegative or all are nonpositive. 
%%%%%%%
It is called  \emph{strictly sign-regular of order~$k$}  (denoted~$SSR_k$)
if it is   sign-regular of order~$k$, and all the minors of order~$k$ are non-zero. In other words, all minors of order~$k$ are non-zero and have the same sign.
For example, if all the entries of~$A$ are nonnegative [positive] then it is~$SR_1$ [$SSR_1$]. 
The  matrix is called 
\emph{sign-regular}~($SR$)
  if it is~$SR_k$ for all~$k$, and
\emph{strictly sign-regular}~($SSR$)
  if it is~$SSR_k$   for all~$k$.
	For example, the matrix~$\begin{bmatrix} 1 & 1/4\\ 40 &2\end{bmatrix}$ is~$SSR_1$ because all its~$1\times 1$ minors are positive, $SSR_2$ because its single~$2\times 2$ minor is negative, and thus it is~$SSR$. 
	
	$SR$ and $SSR$
	 matrices are important in various fields. 
The most prominent examples are totally nonnegative~(TN) [totally positive~(TP)]
 matrices, that is,
matrices  with all minors nonnegative [positive].
Such matrices have beautiful 
 properties and have found
applications in  statistics, computer graphics, approximation theory,
 and more~\citep{total_book,pinkus,gk_book,fallat2017total}.
 
A very important property  of~TN and~TP matrices is  that 
multiplying a vector by such a matrix can only decrease the number of sign variations (see, e.g.,~\cite[Chapter 1]{total_book}). 
This is known as the \emph{variation diminishing property}~(VDP).
Specifically, 
 if~$A\in\R^{ n\times m} $ is~TN
	then
	\[
					s^-(Ax)\leq s^{-}(x) 
	\text{ for all } x\in\R^m,
	\]
	and if~$A$ is~TP   then
\[
					s^+(Ax)\leq s^-(x) 
	\text{ for all } x\in\R^m \setminus\{0\}.
	\]

There is a renewed interest in such VDPs in the context of dynamical 
 systems.  
  \cite{margaliot2019revisiting}   showed
 that 
strong results on the asymptotic behavior of
 nonlinear time-varying tridiagonal cooperative dynamical systems derived by~\cite{smillie} 
 and~\cite{periodic_tridi_smith} 
can be derived using the fact that the transition matrix~$\Phi(t,t_0)$ corresponding 
to  their variational system 
is~TP for all~$t>t_0$ (see also~\cite{Eyal_Smiliie}). In other words, the
variational system
is a    
\emph{totally positive differential system}~(TPDS)~\citep{schwarz1970}.
These transition matrices are real, square, and non-singular. 
Another  recent paper showed that the   transition matrix satisfies a VDP 
with respect to the \emph{cyclic} number of sign variations iff 
 it is~$SSR_k$ for all \emph{odd}~$k$~\citep{CTPDS}.
 \cite{rola_spect} studied the spectral properties of matrices that are~$SSR_k$ for some order~$k$ and introduced the notion of a totally positive discrete-time system. 
\cite{rami_osci} recently generalized this to the notion of  an
oscillatory discrete-time system.  

The next result   describes the equivalence 
 between~$SSR_k$ 
  and a special kind of VDP. 
%%%%%%
\begin{Theorem}\label{thm:gtre}~\citep{CTPDS}
Let~$A\in\R^{n\times n}$ be a nonsingular matrix. 
  Pick~$k \in [1,n]$. 
Then the following  two conditions are equivalent:
%%%%%%%%%%%%
\begin{enumerate}[(a)]
\item	\label{cond:frty} For any vector~$x\in\R^n\setminus\{0\}$ with~$ s^-(x) \leq k-1$, we have~$s^+(Ax)\leq k-1$.

\item			\label{cond:sftry}
$
A 
$
is~$SSR_{k}$. 
\end{enumerate}
\end{Theorem}

\begin{Example}
%%%%%%%%%%%%%%%%%%
For the particular case~$k=1$ Thm.~\ref{thm:gtre}
implies that for a 
 nonsingular matrix~$A\in\R^{n\times n}$ the following properties are equivalent:
\begin{enumerate}[(a)]
\item For any~$x\in\R^n\setminus\{0\}$ with~$ s^-(x) =0$
the entries of~$Ax $ are either all positive or all negative; 
 %%%%%
\item	 The entries of~$A$ are either  all positive or all negative. 		 
\end{enumerate}
\end{Example}
%%%%%%

Note that Thm.~\ref{thm:gtre} does not imply in general that~$s^+(Ax)\leq s^-(x)$. 
However if~$A$ is square and TP (and thus nonsingular)
 then Condition~\eqref{cond:sftry} holds for any~$k$ and this implies the following. 
Pick~$x\in\R^n\setminus\{0\}$, and  let~$k$ be such that~$s^-(x)=k-1$. Then~$s^+(Ax)\leq k-1$, i.e.,~$s^+(Ax)\leq s^-(x)$
and this recovers the~VDP of (square) TP matrices.

For our purposes below, we  also  need the next result that states an 
analogue of Thm.~\ref{thm:gtre} for~$SR_k$
 matrices. 
%%%%%%%%%%%%%%%%%%%%%%%%%%%%%%%
\begin{Theorem}\label{thm:srkew}
Let~$A\in\R^{n\times n}$ be a nonsingular matrix. 
  Pick~$k \in [1,n]$. 
Then the following  two conditions are equivalent:
%%%%%%%%%%%%
\begin{enumerate}[(a)]
\item	\label{cond:onedip} 
For any vector~$x\in\R^n $ with~$ s^-(x) \leq k-1$, we have 
\be\label{eq:dcr}
%%%%%
					s^-(Ax)\leq k-1.
%%%%
\ee
\item			\label{cond:secdip}
$
A 
$
is~$SR_{k}$. 
\end{enumerate}
\end{Theorem}
\noindent The proof follows from a standard 
continuity argument and is given, for the sake of completeness, in the Appendix.

For example,  for the particular case~$k=1$ this implies that for a   nonsingular matrix~$A\in\R^{n\times n}$ the following properties are equivalent:
\begin{enumerate}[(a)]
\item For any~$x\in\R^n $ with~$ s^-(x) =0$
the entries of~$Ax $ are either all nonpositive or all nonnegative; 
 %%%%%
\item	 The entries of~$A$ are either all nonpositive or all nonnegative. 		 
\end{enumerate}

\begin{Remark}
%%%%%%%%%%%%%%%%%%%%%%%%%%%%%
Recall that a vector~$x\in\R^n$ is called
 \emph{totally nonzero} if~$x_i\not =0$ for all~$i\in[1,n]$.
Let~$\text{TNV}_k$ denote the set of all totally  
nonzero vectors~$x\in\R^n$ with~$\sigma(x)=k$ 
(and then of course~$s^-(x)=s^+(x)=k$ as well). 
%%%
\cite{sign_pres_mats} studied the set of nonsingular matrices that map~$\text{TNV}_k$ 
to itself. However, these matrices are quite different from the ones studied
 in this paper,
due to the requirement that every entry of~$Ax$ must be nonzero.
%%%%
\end{Remark}

Another important property of TN matrices, that   will be used
 below to analyze the geometry of the invariant sets of~$k$-positive systems,
 is their spectral structure. All the eigenvalues of a~TN matrix 
  are real and nonnegative, and the corresponding eigenvectors
have special sign patterns. 
A matrix~$A\in\R^{n\times n}$ is called \emph{oscillatory}
if it is~TN and
there exists an integer~$k\geq 1$ such that~$A^k$ is~TP~\citep{gk_book}.
The special spectral structure 
  is particularly evident in the case of oscillatory matrices.
%%%%
\begin{Theorem}\label{thm:spec}\citep{gk_book,Pinkus1996}
%%%%%%%%%%
If~$A\in\R^{n\times n}$ is an oscillatory matrix then its eigenvalues are all real, positive, and distinct.
Order the eigenvalues as~$\lambda_1>\lambda_2>\dots >\lambda_n>0$,
and let~$u^k\in\R^n $ denote the eigenvector corresponding to~$\lambda_k$.
Then
for any~$1\leq i \leq j \leq n$ and any real scalars~$c_i,\dots,c_j  $, that are not all zero,
\be\label{eq:svano}
			i-1 \leq s^-( \sum_{k=i}^j c_k u^k ) \leq s^+ (\sum_{k=i}^j c_k u^k )  \leq   j-1.
\ee
\end{Theorem} 
Note that this implies in  particular that
$			s^-(u^i)=s^+(u^i)= i-1$  for all~$i\in[1,n]$. 
 
	\begin{Example}\label{exa:spec}
		%%%
		Consider the   oscillatory matrix
		$A =\begin{bmatrix}   2&1&0 \\ 1&3&1 \\0 &1 &2   \end{bmatrix} 
		$. Its eigenvalues are~$\lambda_1=4$, $\lambda_2=2$, $\lambda_3=1$, 		
		with corresponding eigenvectors 
	$u^1=\begin{bmatrix} 1 & {2} & 1\end{bmatrix}' $,
	$u^2=\begin{bmatrix} -1 & 0 & 1\end{bmatrix}' $,
	and $u^3=\begin{bmatrix} 1 & -1 & 1\end{bmatrix}' $. Note that~$s^-(u^k)=s^+(u^k)= k-1$ for all~$k\in[1,3]$. 
		%%%
		\end{Example}

In the context of dynamical systems, 
the question is not when does a static mapping satisfy a~VDP,
but rather when does the transition matrix of the system satisfies
 a VDP for all time.  
%%%%%%%%%%%%%%%%%
As shown by~\cite{schwarz1970}, this can be   
analyzed using the 
  dynamics  of \emph{compound 
 matrices}~\citep{muldo1990}.

%%%%%%%%%%%%%%%%%%%%%%%%%%%%%%%%%%%%%%%%%%%%
\subsection{Compound matrices}\label{subsec:compound}
%%%%%%%%%%%%%%%%%%%%
 
Given~$A\in\R^{n\times n}$ and~$k\in[1,n]$, consider the
$\binom{n}{k}^2$
 minors of  order~$k$ of~$A$. 
Each minor is defined by a set of row indices~$1\leq i_1<i_2<\dots<i_k\leq n$ and column indices~$1\leq j_1<j_2<\dots<j_k\leq n$. This minor 
is denoted by~$A(\alpha|\beta)$, where~$\alpha:=\{i_1,\dots,i_k\}$ and~$\beta:=\{j_1,\dots,j_k\}$.
With  a slight abuse of notation we will sometimes treat such ordered sequences as sets. 
For example, for~$A=\begin{bmatrix} 4&5 &6 \\ -1 &4 &-2 \\0&3&-3
\end{bmatrix}$,~$\alpha=\{ 1,3\} $, and~$\beta=\{2,3\}$, we have
\[
A(\alpha|\beta)=\det  \begin{bmatrix} 5&6\\3&-3
\end{bmatrix}   =-33.
\]

For~$A\in\R^{n\times n}$ and~$k\in[1,n]$ 
the~$k$'th \emph{multiplicative  compound matrix}~$A^{(k)}$ 
of~$A$ is the~$\binom{n}{k}\times  \binom{n}{k}$ matrix
that 
includes all these minors ordered lexicographically. 
For example, for~$n=3$  and~$k=2$,    $A^{(2)}$ is the~$3\times 3$ matrix
\[
%%%
%%A^{(2)}=
\begin{bmatrix}
						A(\{1,2\}|\{1,2\}) & A(\{1,2\}|\{1,3\}) & A(\{1,2\}|\{2,3\})\\
						A(\{1,3\}|\{1,2\}) & A(\{1,3\}|\{1,3\}) & A(\{1,3\}|\{2,3\})\\
						A(\{2,3\}|\{1,2\}) & A(\{2,3\}|\{1,3\}) & A(\{2,3\}|\{2,3\}) 
\end{bmatrix}.
%%%
\] 
%%%%%%%%%%%%%%%%%%%%%%%%%%%%%%%%%%%%%%%%%%%%%%
Note that~$A^{(1)}=A$ and~$A^{(n)}=\det(A)$.

\begin{Remark}
A matrix~$A$ is~$SR_k$ 
 iff all the entries
 of~$A^{(k)}$ are either 
all  nonnegative or all nonpositive. In the first  
case~$A^{(k)}$ maps the cone~$\R^{\binom{n}{k}}_+$  
 to itself.
 \cite{kushel2012cone}
studied matrices~$A$ such that for any~$k$
the matrix~$A^{(k)}$ preserves a proper cone. 
%%%%
\end{Remark}

The Cauchy-Binet formula (see, e.g.,~\cite[Ch.~1]{total_book}) asserts  
that 
$
%%%
(AB)^{(k)}=A^{(k)} B^{(k)}.
$
This justifies the term multiplicative compound. 

The $k$'th \emph{additive compound matrix} of~$A$
is   defined  by
\[
				A^{[k]}:= \frac{d}{d \varepsilon}  (I+\varepsilon A)^{(k)} |_{\varepsilon=0}.
\]
This implies that
\be\label{eq:poyrt}
(I+\varepsilon A)^{(k)}= I+\varepsilon   A^{[k]} +o(\varepsilon )  .
\ee 
%%%%%%
%%%
%% 

\begin{Example}
%%%
Consider the case~$n=3$ and~$k=2$. Then
\begin{align*}
						&(I+\varepsilon A)^{(2)}   =\begin{bmatrix}
						1+\varepsilon a_{11} & \varepsilon a_{12} & \varepsilon a_{13}\\ 
						\varepsilon a_{21} &1+\varepsilon a_{22} & \varepsilon a_{23}\\
						\varepsilon a_{31}&\varepsilon a_{32} & 1+\varepsilon a_{33}
						\end{bmatrix}^{(2)}\\
						%%%%%%%%%%%%%%%%%%%%%%%%%%%%%%%%%%%
						&=\begin{bmatrix}
						1+\varepsilon (a_{11}+a_{22}) & \varepsilon a_{23} & -\varepsilon a_{13}\\
						%%%%%%
						\varepsilon a_{32} &1+\varepsilon( a_{11}+a_{33}) & \varepsilon a_{12}\\
						%%%%%%
						-\varepsilon a_{31}&\varepsilon a_{21} & 1+\varepsilon (a_{22}+a_{33})
						\end{bmatrix}+o(\varepsilon),
\end{align*}
%%%
so 
\begin{align}\label{eq:case32}
							A^{[2]}&=
						\frac{d}{d \varepsilon}  (I+\varepsilon A)^{(2)} |_{\varepsilon=0}\nonumber \\&=
						\begin{bmatrix}
						 a_{11}+a_{22} &   a_{23} & -  a_{13}\\
						%%%%%%
						  a_{32} &  a_{11}+a_{33} &  a_{12}\\
						%%%%%%
						-  a_{31}&  a_{21} &  a_{22}+a_{33}
						\end{bmatrix}.
\end{align}
%%%
\end{Example}

The  Cauchy-Binet  formula can be used to prove that~$
(A+B)^{[k]}= A^{[k]}+B^{[k]},
$
thus justifying the term additive compound.

The additive compound arises naturally 
 when studying the dynamics of the multiplicative compound. 
For a time-varying matrix~$Y(t)$ let~$Y^{(k)} (t) := (Y(t))^{(k)} $. 
Suppose that~$Y(t)$ evolves according to~$\frac{d}{dt} Y(t)=A(t)Y(t)$. 
Then a Taylor approximation yields 
\begin{align*}
      Y^{(k)} (t+\varepsilon)&=  (Y(t)+\varepsilon A(t)Y(t)   ) ^{(k)}+o(\varepsilon)\\
			&=(I+\varepsilon A(t)   ) ^{(k)} Y  ^{(k)}(t)+o(\varepsilon),
\end{align*}
and combining this with~\eqref{eq:poyrt} gives
\be\label{eq:povt}
\frac{d}{dt} Y^{(k)}(t)=A^{[k]}(t) Y^{(k)}(t),
\ee
where~$ A^{[k]}(t):=(A(t)) ^{[k]}$.
Thus, the dynamics of all the minors of order~$k$ of~$Y(t)$, stacked in the matrix~$Y^{(k)}(t)$, is also described by a  linear dynamical  system,  
  with the matrix~$A^{[k]}(t) $.
	
For any~$k\in[1,n]$, the matrix~$A^{[k]}$ can be given
 explicitly in terms of the entries~$a_{ij}$ of~$A$. 
\begin{Lemma}\label{lem:poltr}
%%%%%%%%%%%%%%%%%%%%%%%%%%%%
The entry of~$A^{[k]}$ corresponding to~$(\alpha|\beta)=(i_1,\dots,i_k|j_1,\dots,j_k) $  is:
\begin{itemize}
%%%
\item $\sum_{\ell=1}^k a_{i_\ell i_\ell}$ if~$i_\ell=j_\ell$ for all~$\ell\in[1,k]$;
%%%%% 
\item $(-1)^{\ell+m} a_{i_\ell j_m}$ 	 if all the indices in~$\alpha$ and~$\beta$ coincide except for a single index~$i_\ell\not = j_m $; and
 %%%%
\item $0$, otherwise. 
\end{itemize}
\end{Lemma} 
%%%%
\noindent For a proof of this result, see e.g.,~\cite{schwarz1970}
or~\cite{fiedler_book}.

The first case in Lemma~\ref{lem:poltr}
   corresponds to diagonal entries  of~$A^{[k]}$.  
All the other entries of~$A^{[k]}$ are
 either zero or     an entry of~$A$
multiplied  by either plus or minus one.

\begin{Example}
%%%%%%%%%%%%%%%%%%%%%%%%%%%%%%%%%%%
 Consider the case~$n=4$, i.e.,~$A=\{a_{ij}\}_{i,j=1}^4$. 
Then Lemma~\ref{lem:poltr} yields
%%%%
 %%%%%%%%%%
\begin{align}\label{eq:a24}
%%%%%
A^{[2]}&=\left [ \begin{smallmatrix} 
																%%%
																 a_{11}+a_{22}   & a_{23} & a_{24} & -a_{13} & -a_{14} & 0 \\
																%%%%%
																a_{32}      &a_{11}+a_{33} & a_{34} & a_{12} & 0 & -a_{14} \\   
																%%%%%
																a_{42}     &a_{43}   &a_{11}+a_{44} & 0 & a_{12} & a_{13} \\   
																%%%%%
																-a_{31}      &a_{21} & 0 & a_{22}+a_{33} & a_{34} &  -a_{24} \\
																	%%%%%
																-a_{41}      &0 & a_{21} & a_{43} & a_{22}+a_{44} &  a_{23} \\ 
																	%%%%%
																0      &-a_{41} & a_{31} & -a_{42} & a_{32} &  a_{33}+a_{44}  
%%%%%%%%%%%%
  \end{smallmatrix}\right] , 
	\end{align}
and
\begin{align*}
%%%%%
A^{[3]}&=\left[ \begin{smallmatrix}     
													%%%
																a_{11}+a_{22}+a_{33} & a_{34} & -a_{24} & a_{14} \\
																%%%%%			
																a_{43}& a_{11}+a_{22}+a_{44} & a_{23} & -a_{13}  \\
																%%%%%			
																-a_{42}& a_{32} & a_{11}+a_{33}+a_{44} & a_{12}  \\
																%%%%%			
                                a_{41}& -a_{31}& a_{21}& a_{22}+a_{33}+a_{44} 
																%%%%%			
                          %%%%
 \end{smallmatrix} \right]. 
\end{align*}
The entry in the first row and third column of~$A^{[3]}$
corresponds to~$(\alpha|\beta)=( \{1,2,3\}| \{1,3,4\})$, and since~$\alpha$ and~$\beta$ coincide except for the entry~$\alpha_{i_2}=2$   and~$\beta_{j_3}=4$, this entry is~$(-1)^{2+3} a_{i_2 j_3} = -a_{24}$.
It is useful to index compound matrices using~$\alpha,\beta$.
For example,  we write
\[
A^{[3]}( \{1,2,3\}| \{1,3,4\})= -a_{24}.
\]
%%%
\end{Example}

We note  two  
 special cases of~\eqref{eq:povt}. 
For~$k=1$,~$Y^{(1)} $ is the matrix that contains the first-order minors of~$Y$, that is,~$Y^{(1)}=Y$, and Lemma~\ref{lem:poltr} gives~$A^{[1]}=A$, 
so~\eqref{eq:povt} becomes~$\dot Y=AY$. For~$k=n$, 
$Y^{(n)}$ is the matrix that contains all the~$n\times n$ minors of~$Y$, that is,~$\det Y$, and using Lemma~\ref{lem:poltr}  yields
\[
\frac{d}{dt} (\det Y(t))=\tr (A(t))  \det Y(t),
\]
which is the   Abel-Jacobi-Liouville identity (see, e.g.~\cite{Byrnes_global}).

For our purposes, it is important to determine whether for a given~$A\in\R^{n\times n}$
the matrix~$A^{[k]}$ 
is Metzler or not. This can be done using Lemma~\ref{lem:poltr}.
The next result demonstrates this.
We require the following definition.
\begin{Definition}\label{def:kset}
%%%%% 
Let~$M^n_2 $
denote the set of matrices~$A\in\R^{n\times n}$ satisfying:
\begin{enumerate}[(a)]
\item $a_{1n},a_{n1} \leq 0$; 
\item $a_{ij}\geq 0$ for all~$i,j$ with~$|i-j|=1$;
\item $a_{ij} =  0$ for all~$i,j$ with~$1<|i-j|<n-1$.
\end{enumerate}
\end{Definition}
For example, for~$n=5$
the matrices in~$M^5_2$ are those with the sign pattern
\[
\begin{bmatrix}
*       & \geq 0 & 0 & 0& \leq 0 \\
 \geq 0 & * & \geq 0 &  0&  0 \\
 0& \geq 0 & * & \geq 0 &  0 \\
 0& 0& \geq 0 & * & \geq 0  \\
 \leq 0 & 0&  0 & \geq 0 & *  
\end{bmatrix},
\]
where~$*$ denotes ``don't care''. 

%%%%
\begin{Lemma}\label{lem:a[22]}
%%%%%%%%%%%%%%%%%%
Let~$A \in \R^{n\times n  }$ with~$n>2$. Then~$A^{[2]}$ is Metzler iff~$A\in M^n_2$.
%%%%
\end{Lemma}
\begin{Example}
Consider the case~$n=4$. In this case~$A^{[2]}$ is given in~\eqref{eq:a24}
and it is straightforward to verify that~$A^{[2]}$  is Metzler iff
$a_{12},a_{23},a_{34},a_{21},a_{32},a_{43} \geq 0$, 
$a_{13}=a_{24}=a_{31}=a_{42}=0$, and~$a_{14},a_{41}\leq 0$,
that is, iff~$A \in M^4_2$.
\end{Example}

{\sl Proof of Lemma~\ref{lem:a[22]}.}
%%%%%%%%%%%%%%%%%%%%%%%%%%%%%%%%%%%%%%
It follows from Lemma~\ref{lem:poltr} 
that for any~$i\not = j$
   the
 entry~$  a_{ij}$  or~$(-a_{ij})$ appears as an
offdiagonal entry of~$A^{[2]}$ iff 
one of  the following cases holds for some~$p\in[1,n]$:
%%%%%%%%%%%%
\begin{enumerate}[(1)]
%%%%
\item if~$i < p< j $   then~$A^{[2]}(\{ i , p\} | \{p, j \})=-a_{ij} $;
\item  if~$p<i$ and~$p<j$
  then~$A^{[2]} (\{ p,i\} | \{p, j \}) = a_{ij}$;
%%%%
\item  if~$p>i$ and~$p>j$
  then~$A^{[2]}(\{ i,p\} | \{j,p \}) =  a_{ij}$;
%%%%%%%%%%%
\item  if~$j< p<i$ 
 then~$A^{[2]}(\{ p,i\} | \{j,p \}) = - a_{ij}$.
%%%%%
\end{enumerate} 

Consider the case~$i=1$ and~$j=n$. Then only case~(1) applies and we conclude
 that~$-a_{1n}$ (but not~$a_{1n}$)
appears in~$A^{[2]}$, so if~$a_{1n}>0$ then~$A^{[2]}$ is not Metzler.
A similar argument using case~(4) shows that~$-a_{n1}$
 appears in~$A^{[2]}$, so if~$a_{n1}>0$ then~$A^{[2]}$ is not Metzler.

Pick~$i,j\in[1,n]$  with~$|i-j|=1$. 
Then cases~(1) and~(4) do not apply, whereas cases~(2) and~(3) 
imply that~$a_{ij}$ appears in~$A^{[2]}$. This entry must be 
nonnegative, or else~$A^{[2]}$ is not Metzler.

Pick~$i,j\in[1,n]$ with~$1<|i-j|<n-1$. 
Then it can be shown using cases~(1)-(4)
that both~$a_{ij}$ and~$-a_{ij}$ 
appear in~$A^{[2]}$ and thus if~$a_{ij}\not =0$ then~$A^{[2]}$ is not Metzler. 
We conclude that if~$A\not \in M^n_2$ then~$A^{[2]}$ is not Metzler.
But the arguments above also show that if~$A  \in M^n_2$ then~$A^{[2]}$ is   Metzler.
This completes the proof of Lemma~~\ref{lem:a[22]}.~\hfill{$\square$}

Let~$\M \subset \R^{n\times n}$ [$\M^+ \subset \R^{n\times n}$] denote the set of matrices 
that are tridiagonal, and   with nonnegative [positive] 
entries on the super- and sub-diagonals. One implication of
 Lemma~\ref{lem:a[22]}
is that~$A^{[1]} =A$ and~$A^{[2]}$ are both
  Metzler iff~$A\in \M$. 
If, in addition, we require~$A$ to be irreducible then this holds
iff~$A\in\M^+$~\citep{margaliot2019revisiting}.  \cite{schwarz1970}
showed that the transition matrix~$\exp(At)$  
 is~TP for all~$t>0$ iff~$A \in\M^+$.

 We are now ready to define a generalization of a positive~LTV system.
%%%%%%%%%%%%%%%%%%%%%%%%%%%%%%%%%%%%%%%%%%%%%%%%%%
\section{$k$-positive linear systems}\label{sec:def_k_pos}
%%%%%%%%%%%%%%%%%%%%%%%%%%%%%%%%%%%%%%%%%%%%%%%%%%%%
For any~$k\in[1,n]$,
define the sets
\[
     P^k_-:=\{ z\in\R^n : s^-(z)\leq k-1 \},
\]
and
\[
     P_+^k:=\{ z\in\R^n : s^+(z)\leq k-1 \}.
\]
It is not difficult to show  that~$P^k_-$ is closed,
$P^k_+$ is open.
Note that
\be\label{eq:p1}
P^1_-=\R^n_+ \cup \R^n_- , \quad  P^1_+=\Int \R^n_+ \cup  \Int \R^n_- ,
\ee
and  that 
\begin{align}\label{eq:psdv}
%%%
&	 P_+^k=\Int( P^k_-) \text{ for all } k\in[1,n-1],\nonumber \\ 
%%%
&   P^1_- \subset P^2_- \subset \dots \subset P^n_-=\R^n,\nonumber\\
&   P^1_+ \subset P^2_+ \subset \dots \subset P^n_+=\R^n.
\end{align}

\begin{Remark}

	Several authors studied related sets of vectors.
	 \cite{oliva1993diffeomorphisms} studied 
	diffeomorphisms~$f:\R^n\to\R^n$ whose Jacobian~$J(x)$
	is an oscillatory matrix for all~$x\in\R^n$,   
	and defined sets that are closely related to~$P^k_-$ and~$P^k_+$. 
	In Section~\ref{sec:geom} below we analyze the geometrical structure of~$P^k_-$,
	and in particular show that they are  cones of rank~$k$
	(see also~\cite[Ch.~1]{pls_sobolev}).
	
%%%%%%%%%%%%%%%%%%%%%%%%%%%%%
 \end{Remark}

Fix a time interval~$-\infty\leq a<b\leq\infty$.
Consider the time-varying linear system:
\be\label{eq:linsysp}
\dot x(t)=A(t)x(t), \quad x(t_0)=x_0,
\ee
where~$A(\cdot):(a,b)\to \R^{n\times n}$ is
a    locally  (essentially)  
bounded measurable matrix function and~$t_0 \in (a,b)$. It is well-known that 
this implies that~\eqref{eq:linsysp}
admits a unique absolutely-continuous 
solution~\citep{sontag_book}.
 This solution satisfies~$x(t)=\Phi(t,t_0)x(t_0)$, where~$\Phi(t,t_0)$ 
 (sometimes written~$\Phi(t)$ for brevity) is the solution at time~$t$ of the matrix
differential equation:
\be\label{eq:mdfs}
\dot \Phi(s  )=A(s)\Phi(s ),\quad \Phi(t_0 )=I, 
\ee

 We are now ready to define the main notion studied in this paper.

\begin{Definition} \label{def:k_pos}
%%%%%%%%%%%%%%%%%
Fix~$k\in[1,n]$.
 We say that~\eqref{eq:linsysp} is \emph{$k$-positive} on the time interval~$(a,b)$
  if~$P^k_-$ 
is an invariant set of the dynamics, that is,
for any pair~$a<t_0<t<b$ and any~$x(t_0)\in P^k_-$ we have~$x(t) \in P^k_-$.
%%%%%%% 
 \end{Definition}
%%%%
%%%%
 \noindent  Eq.~\eqref{eq:p1} implies that a~$1$-positive system is  a positive system.

The next result provides a necessary and sufficient condition for~\eqref{eq:linsysp}  to be~$k$-positive in terms of the~$k$'th additive compound~$A^{[k]}(t)$.
\begin{Theorem}\label{thm:kiff}
The   system~\eqref{eq:linsysp}  is~$k$-positive on~$(a,b)$
 iff~$A ^{[k]}(s)$ 
 is   Metzler   for almost
all~$s\in(a,b)$.
\end{Theorem}
%%%
\emph{Proof of Thm.~\ref{thm:kiff}.}
%%%%%%%
Thm.~\ref{thm:srkew}
 implies that~$k$-positivity is equivalent  to
$\Phi(t,t_0)$ being~$SR_k$ for all~$a<t_0<t<b$, that is, 
either~$\Phi^{(k)}(t,t_0)\geq 0$ or~$\Phi^{(k)}(t,t_0)\leq 0$
for all~$a<t_0<t<b$. 
%%%
  By~\eqref{eq:povt} and~\eqref{eq:mdfs},
	%%%%%%%
\be\label{eq:affs}
			\frac{d}{ds} \Phi^{(k)}(s)=A ^{[k]}(s)\Phi^{(k)}(s),\quad \Phi^{(k)}(t_0)=I.
\ee
By  continuity, this  implies  that~$\Phi(t,t_0)$ 
is~$SR_k$ for all~$a<t_0<t<b$ iff
\be\label{eq:odcx12}
\Phi  ^{(k)} (t,t_0) \geq 0 \text{ for all } a<t_0<t<b.
\ee
It is well-known (see e.g.,~\cite[Lemma~2]{margaliot2019revisiting}) 
that the solution of~\eqref{eq:affs} satisfies~\eqref{eq:odcx12}
    iff~$A^{[k]}(s)$ is Metzler for almost all~$s\in(a,b)$.~\hfill{$\square$}
 %%%%%%%%%%%%%%%

\begin{figure*}[t]
 \begin{center}
%%%%%%%%%%%%%%%%%%%%%%%[width=12cm,height=12cm]
   \includegraphics[scale=0.6]{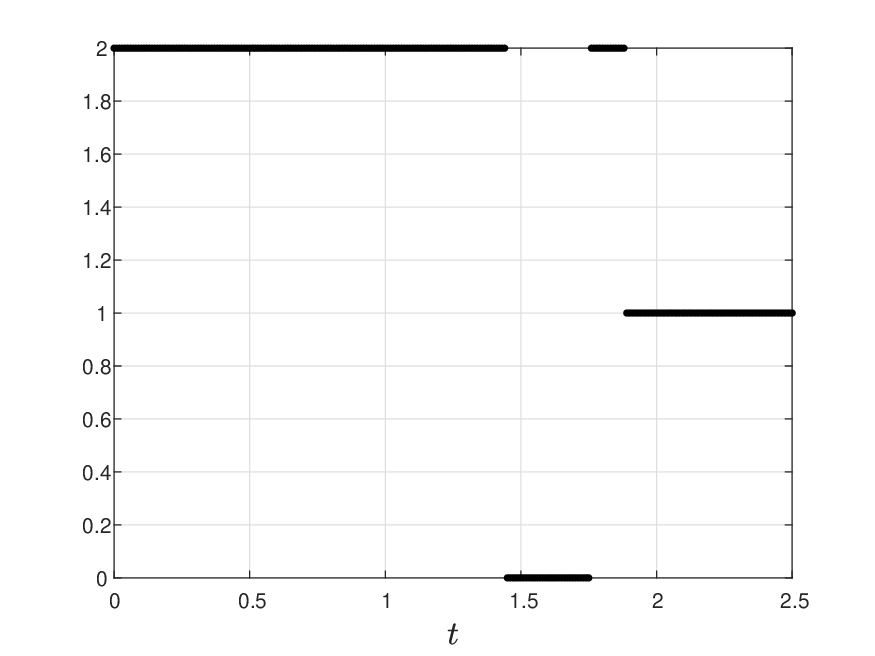}
\caption{ $s^-(x(t))$ as a function of~$t$
for the trajectory~$x(t)$ in Example~\ref{exa:sminuin}. }\label{fig:s2inv}
%%%%%%%%%%%%%%%%%%%%%%%%%%
\end{center}
\end{figure*}

 \begin{Example}\label{exa:sminuin}
%%%
Consider~\eqref{eq:linsysp} with  the constant   matrix
%%%
\[
A  = \begin{bmatrix}     
													%%%
															 -1&2&-2&1 \\
																%%%%%			
															3&0&1&-1 \\
																%%%%%			
								-4&1.5&2&4\\
																%%%%%			
1&-1&2&5																%%%%%			
                          %%%%
 \end{bmatrix}. 
\]
Using   Lemma~\ref{lem:poltr} yields
\begin{align*}
%%%%%%%%%%%%%%%%%%%
A^{[3]}&=\left[ \begin{matrix}     
													%%%
																1 & 4& 1 & 1 \\
																%%%%%			
																2& 4 & 1 & 2  \\
																%%%%%			
																1& 1.5 & 6 & 2  \\
																%%%%%			
                                1& 4& 3& 7 
																%%%%%			
                          %%%%
 \end{matrix} \right],
\end{align*}
%%%%%%%%%%%%%%%%%%%%%%%%%%%%%%%%%%%%%
so~$A^{[3]}$ is Metzler.
 Hence, the system is~$3$-positive, and the set~$P^3_-=\{x\in\R^4:s^-(x)\leq 2\}$ is an invariant set of the dynamics. 
Fig.~\ref{fig:s2inv}
 depicts~$s^-(x(t))$, with~$x(0)=\begin{bmatrix}   
0.34 &   -0.54&   -1.06&    0.49   \end{bmatrix}' $, for~$t\in[0,2.5]$. 
Note that~$s^-(x(0))=2$. 
It may be seen that~$s^-(x(t))$ both decreases and increases yet, as expected,~$s^-(x(t))\leq 2$ for all~$t\geq  0$. 
%%%
\end{Example}

For a given~$A$, the additive compounds~$A^{[1]} ,  \dots, A^{[n]}$  are \updt{related}.
In particular,  \cite{schwarz1970} showed that if~$A ^{[1]} $ and~$A ^{[2]} $ are
   Metzler   then~$A ^{[k]}$
	  is Metzler for every~$k\in[1,n]$.
Combining this with Definition~\ref{def:k_pos} and Thm.~\ref{thm:kiff} 
yields the following result.
%%%%%%%%%%%%%%%%%%%%%%%%%%%%%%%%%%%%%
\begin{Corollary} \label{cor:1_and_2_implies_all_k}
    If the system~\eqref{eq:linsysp} is~$1$-positive and~$2$-positive
    then it is~$k$-positive for all $k\in[1,n]$. 
\end{Corollary}

We now turn to define a stronger notion of~$k$-positivity.
%%%%%%
\begin{Definition} \label{def:strong_k_pos}
%%%%%%%%%%%%%%%%%
Fix~$k\in[1, n]$. 
%%%%%%% 
We say that~\eqref{eq:linsysp} is \emph{strongly $k$-positive} on~$(a,b)$
  if for any pair~$(t_0,t)$ s.t.~$a<t_0<t<b$ we have 
	\[ x(t_0)\in P^k_-\setminus\{0\} \implies 
x(t) \in P^k_+.
\]
\end{Definition}

In other words, the dynamics map~$P^k_-\setminus\{0\}$ to~$P^k_+$.

To provide a sufficient condition for 
strongly~$k$-positivity, we recall one possible  definition for   irreducibility 
of a measurable matrix function~\citep{Walter1997}.
 Let~$J:=(a,b)$.
 A measurable 
set~$M\subset  J$ is said to be \emph{dense at~$a$}
 if the set $M \cap [a, a + \varepsilon ]$ has positive
measure for every~$\varepsilon > 0$.
%%%%
 For measurable 
functions~$f,g: J\to\R$ and~$a\in J$,
we write
$f > g$ at~$a^+$ if the 
set~$\{t \in J | f(t) > g(t)\}$ is dense at~$a$.
%%%%%%%%%%%%%%%%%%%%%%%%%
A measurable matrix function~$C:J\to\R^{n\times n}$
is said to be \emph{irreducible at~$ a^+$}
if for every two nonempty index sets~$\alpha,\beta \subset \{1,\dots, n\}$, 
with~$\alpha \cup \beta =\{1,\dots, n\}$,
and~$\alpha \cap \beta=\emptyset$, 
 there exist
 indices~$k \in \alpha$, $j \in \beta$ such that~$c_{jk}  > 0$
at~$a^+$.

The next result provides a sufficient condition for
strongly~$k$-positivity.

\begin{Theorem}\label{thm:skihh}
Suppose that~$A ^{[k]}(s)$ 
 is   Metzler   for almost
all~$s\in(a,b)$, and  that  for any~$a< t_0<t<b$ 
there exists~$t_0\leq \tau<t$ such that~$A^{[k]}(s)$ is irreducible at~$\tau^+$.
Then~\eqref{eq:linsysp}  is strongly~$k$-positive on~$(a,b)$.
\end{Theorem}
%%%
\emph{Proof of Thm.~\ref{thm:skihh}.}
%%%%%%%
It is well-known~\citep{Walter1997} that the assumptions in the 
statement of the theorem imply that for any~$a<t_0<t<b$ the solution of~\eqref{eq:affs}
 satisfies
\[
\Phi^{(k)}(t,t_0)\gg 0 .
\]
In particular,~$\Phi (t,t_0)$ is~$SSR_k$. 
Pick~$x(t_0)\in P^k_-\setminus \{0\}$.
Then Thm.~\ref{thm:gtre} implies that
\begin{align*}
s^+(x(t))&=s^+(\Phi(t,t_0)x(t_0))\\
&\leq k-1,
\end{align*} 
so~$x(t) \in  P^k_+$.~\hfill{$\square$}
 %%%%%%%%%%%%%%%

For the case where~$A(t)$ is continuous in~$t$  it is possible to 
give a necessary and sufficient condition for strongly~$k$-positivity. 
\begin{Theorem}\label{thm:strict_kiff_co}
Let~$A(\cdot):(a,b)\to\R^{n\times n}$ be  a continuous matrix function.
The system~\eqref{eq:linsysp}  is 
strongly~$k$-positive on~$(a,b)$
 iff  the following two conditions hold:
$A^{[k]}(\tau)$   is Metzler for all~$\tau\in(a,b)$, and for 
 any interval~$[p,q]$, with~$a<p<q<b$, 
there exists~$t^* \in [p,q]$ such that~$A^{[k]}(t^*)$ is   irreducible.
\end{Theorem}

%%%%%%% 
\begin{proof}
%%%%%%% 
 \cite[Lemma 2]{CTPDS} shows that the conditions above are equivalent
to the condition
\[
\Phi  ^{(k)} (t,t_0) \gg 0 \text{ for all } a<t_0<t<b.
\]
Combining this with 
Thm.~\ref{thm:gtre}  completes the proof. 
 %%%%%%%%%%%%%%%
\end{proof}

Our next goal is to study systems 
that are strongly~$k$-positive for several values of~$k$.
Since we are interested in asymptotic properties, we assume from here on that
the time interval is~$(a,b)=(a,\infty)$.
%%%%%%%%%%%%%%%%%%%%%%%%%%%%%%%%%%%%%%%%%%%%%%%%%%%%%%
\begin{Proposition} \label{prop:k_pos_monot_ser}
%%%%%%%
    Assume  that there exists~$k\in [1,n-1]$
		such that~\eqref{eq:linsysp} is 
		strongly~$i$-positive for all~$i  \leq k  $.
		Then for any~$x(t_0)\in P^k_-\setminus\{0\}$ and any set of times
		$  t_0<t_1<t_2<\dots  $ we have 
			\begin{align}\label{eq:setvp}
		s^-(x(t_0)) &\geq s^+(x(t_1))	\geq 
		s^-(x(t_1)) \geq s^+(x(t_2)) \nonumber \\
		&\geq s^-(x(t_2)) 
			\geq s^+(x(t_3))	 \geq \dots,
	\end{align}
		and no more than~$k-1$ inequalities 
			here are strict. 
 Furthermore, 
there exists a time~$\tau\geq t_0 $ such that 
\be\label{eq:truaio}
				x(t) \in \V \text{ for all } t\geq\tau. 
\ee
		%%%%
    \end{Proposition}
		%%%%%%%%%%%%%%%%%%%%%%%%%%
		Note that~\eqref{eq:setvp} implies that both~$s^-(x(t))$ and~$s^+(x(t))$
		are integer-valued Lyapunov functions as they are bounded below
		(by zero) and 
			non-increasing along  
		  any trajectory~$x(t)$ 
		emanating from~$x(t_0)   \in P^k_-\setminus\{0\}$. 
		
		%%%%%%%%%%%%%%%%%%%%%%
\emph{Proof of Prop.~\ref{prop:k_pos_monot_ser}.}
%%%%%%%%%%%%%%%%%%%%%%%%%%%%%%%%%%%%
  Pick~$x(t_0)   \in P^k_-\setminus\{0\}$. 
	Let~$v:=s^-(x(t_0) )$. Then~$v\leq k-1$ and~$x(t_0)\in P^{v+1}_-\setminus\{0\} $.
	Since the system is strongly~$(v+1)$-positive,~$x(t_1)\in P^{v+1}_+$, that is,
	\[
	 s^+(x(t_1))\leq v=s^-(x(t_0)).
	\]
	 In particular,~$w:=s^-(x(t_1))\leq s^+(x(t_1))\leq v $. 
	Since the system is strongly~$(w+1)$-positive,~$x(t_2)\in P^{w+1}_+$, that is,
		\[
	 s^+(x(t_2))\leq w=s^-(x(t_1)).
	\]
	Continuing in this manner yields~\eqref{eq:setvp}.

Since~$s^-,s^+$  take values in~$ [0,k-1]$,
no more  than~$k-1$ inequalities in~\eqref{eq:setvp} 
			can be    strict. Let~$\tau_i$ denote the (up to~$k-1$) time points where
			$s^+(x(\tau_{\ell+1}))<s^-(x(\tau_\ell))$. Then~\eqref{eq:truaio}
			holds for~$\tau: =\max_i \tau_i$.~\hfill{$\square$}.

Prop.~\ref{prop:k_pos_monot_ser} implies in particular
that if the system
is 
		strongly~$i$-positive for all~$i  \in [1,n-1] $
		then~\eqref{eq:truaio} holds for any~$x(t_0)\not =0$.
		This recovers an important 
		result in~\citep{schwarz1970}, 
		\updt{which states that for a~TPDS Eq.~\eqref{eq:truaio} holds for any~$x(t_0)\not =0$.}

Thm.~\ref{thm:kiff} provides 
  a condition on~$A^{[k]}$ ensuring that the linear  system~\eqref{eq:linsysp} is~$k$-positive.
We now turn to express this condition in terms of~$A$.
%%%%%%%%%%%%%%%%%%%%%%%%%%%%%%%%%%%%%%%%%%%%%%%%%%%%%%%%%%%%%%
\section{Explicit algebraic conditions for $k$-positivity}\label{sec:expa}
%%%%%%%%%%%%%%%%%%%%%%%%%%%%%%%%%%%%%%%%%%%%%%%%%%%%%%%%%%%%%%
We begin by  considering the   case~$k=n-1$. 

\subsection{$(n-1)$-positive systems}
%%%%
Given~$A\in\R^{n\times n}$, when is~$A^{[n-1]}$ Metzler?
To address this question we require the following definition.

%%%%%%%%%%%%%%%%%%%%%%%%%%%%%%%%%
\begin{Definition}\label{def:knm1}
	%%%%% 
	Let~$M^n_{n-1}$ denote the set 
	   of matrices~$A\in\R^{n\times n}$ satisfying 
	$a_{ij}\geq 0$ for all~$i,j$ such that~$i-j $ is odd,
	and~$a_{ij}\leq 0$ for all~$i\not =j$ such that~$i-j $ is even.
	%%%%%%%%%%%%%%
	\end{Definition}
%%%%%%%%%%%%%%%%%%%
For example, for~$n=4$ the matrices in~$M^4_3$ are those with the sign pattern:
\[
\begin{bmatrix}
												*& \geq 0& \leq 0& \geq  0\\
												\geq 0 &* &\geq 0 & \leq 0 \\
												\leq 0& \geq 0&*& \geq 0\\
												\geq 0&\leq 0&   \geq 0&*
\end{bmatrix},
\]
where~$*$ denotes ``don't care''.
In particular,  the matrix~$A$
 in Example~\ref{exa:sminuin} satisfies~$A\in M^4_3$. 

\begin{Lemma}\label{lem:explicit_anm1}
	%%%
	Let~$A \in \R^{n\times n  }$ with~$n>2$. 
	Then~$A^{[n-1]}$ is Metzler iff~$A\in M^n_{n-1}$.
	%%%%%%
\end{Lemma}

{\sl Proof of Lemma~\ref{lem:explicit_anm1}.}
%%%%%%%%%%%%%%%%%%%%%%%%%%%%%%%%%%%%%%%%%%%%%%%%
It follows from Lemma~\ref{lem:poltr} 
that an off-diagonal entry of~$A^{[k]}$ 
corresponding to~$(\alpha|\beta)=(i_1,\dots,i_k|j_1,\dots,j_k) $
can be nonzero  only if  all the indices in~$\alpha$ and~$\beta$ coincide, except for a single index~$i_\ell\not = j_m $, and then  
$A^{[k]}(\alpha,\beta)=(-1)^{\ell+m} a_{i_\ell j_m}$.
We use this to determine when~$a_{pq}$ (or~$-a_{pq}$) appears 
on an off-diagonal entry of~$A^{[n-1]}$.  We   consider only
pairs~$(p,q)$ with~$q\geq p$, as the case~$p\geq q$ follows by symmetry. It is clear that if~$p=q$ then~$a_{pq}=a_{pp}$
does not appear as an off-diagonal entry of~$A^{[n-1]}$. 
%%%%%%
Pick~$p,q$ with~$1\leq p<q\leq n$. 
Suppose that~$a_{pq}$ or~$-a_{pq}$ appears as an off-diagonal entry of~$A^{[n-1]}$ corresponding to~$(\alpha|\beta)$. 
 This implies that
$\alpha=\{i_1,i_2,\dots,i_{n-1}\}$ and
$\beta=\{j_1,j_2,\dots,j_{n-1}\}$
 coincide except  for a single index~$i_\ell\not=j_m$,
with~$i_\ell=p$ and~$j_m=q$. 
Thus,~$\alpha=\{1,2,\dots,n\}\setminus \{q\}$ 
and~$\beta=\{1,2,\dots,n\}\setminus \{p\}$.
Since~$p<q$, this gives~$i_\ell=i_p$ and~$j_m=j_{q-1}$,
so~$A^{[n-1]}(\alpha|\beta)=(-1)^{p+q-1}a_{pq}$.
By symmetry, we conclude that for any~$p\not=q$ 
we have that~$(-1)^{p+q-1}a_{pq}$
is an off-diagonal 
entry of~$A^{[n-1]}$. 	
Hence,~$A^{[n-1]}$
is Metzler iff~$(-1)^{p+q-1}a_{pq} \geq 0$ for all~$p\not =q $.~\hfill{$\square$}

\begin{Remark}
%%%%%%%
For~$F\in\R^{n\times n}$, let~$\tilde F$ denote the matrix with entries
\[
\tilde f_{ij}:=(-1)^{i+j} f_{n+1-i,n+1-j},\quad i,j \in[1,n].
\]
 \cite{schwarz1970} proved that if~$A\in\R^{n\times n}$ then
$
			A^{[n-1]} = \tilde { B     },
$
where~$B:=\tr(A) I -A'$. This implies that~$A^{[n-1]}$ is Metzler iff
$(-1)^{i+j+1} a_{n+1-j,n+1-i}\geq 0$ for all~$i\not =j$. This provides an alternative proof of
 Lemma~\ref{lem:explicit_anm1}.
%%%%%%%%
\end{Remark}

\begin{Example}
%%%%%%
Consider the case~$n=3$ and~$k=n-1=2$. Then
\begin{align*}
B&:=\tr(A) I -A'\\
&=\begin{bmatrix}
									a_{22}+a_{33}  &   -a_{21}    & -a_{31} \\
									-a_{12}  &   a_{11}+a_{33}    & -a_{32} \\
									-a_{13}   &   -a_{23}    & a_{11}+a_{22}   
\end{bmatrix},
\end{align*}
so
\begin{align*}
A^{[2]}&=\tilde B \\
&=\begin{bmatrix}
									a_{11}+a_{22}  &   a_{23}    & -a_{13} \\
									a_{32}  &   a_{11}+a_{33}    & a_{12} \\
									-a_{31}   &   a_{21}    & a_{22}+a_{33}   
\end{bmatrix},
\end{align*}
%%%%%
and this agrees with~\eqref{eq:case32}.
%%%%%%
\end{Example}

Recall that the system~$\dot x=Ax$ is called a \emph{competitive system}
if~$(-A)$ is Metzler~\citep{hlsmith}. 
The next result shows that~$(n-1)$-positive systems are just competitive  systems in disguise. 
%%%%%%
\begin{Lemma}
Let~$D:=\diag(1,-1,1,\dots,(-1)^{n-1})$,
 and let~$P\in\R^{n\times n}$ denote the permutation matrix
\[
P:=\begin{bmatrix}
																	0&0&\dots&0&0&1\\
																	0&0&\dots&0&1&0\\
																	0&0&\dots&1&0&0\\
																	&&\vdots\\
																	1&0&\dots&0&0&0
	%%%%%%%%																
\end{bmatrix}.
\]
Note that~$D^{-1}=D$ and~$P^{-1}=P$.
	%%%
	 Consider the system~$\dot x(t)=Ax(t)$, and let~$y(t):=-DPx(t)$, so that~$
\dot y(t)=B y(t),
$
 with~$B:=DPAPD$. 
		The following two conditions are equivalent.
	\begin{enumerate}
	\item $A\in M^n_{n-1}$, i.e.~$\dot x=Ax$ is~$(n-1)$-positive;
	\item the matrix~$(-B) $ is Metzler, i.e.~$\dot y=B  y$ is competitive.
	%%%%%%
	\end{enumerate} 
	%%%%%%
\end{Lemma}
%%%%%%%%%%%%%%%%%%%%%%%%%%%%%%%
\begin{proof}
%%%%%%%%%%%%%%%%%%%%%%%%%%%%%%%
Let~$C:=PAP$. Then~$c_{ij}=a_{ n+1-i,n+1-j}$.
Since~$-B=-DCD$, 
\begin{align*}
-b_{ij}& = (-1)^{i+j+1} c_{ij}\\
&= (-1)^{i+j+1} a_{ n+1-i,n+1-j}\\
&= (-1)^{ n+1 + i-(n+1-j)    +1 } a_{ n+1-i,n+1-j},
%%%%%%%%%%%
\end{align*}  
and the  definition of~$M^n_{n-1}$ implies that~$-b_{ij}\geq 0$ for all~$i\not =j$
iff~$A\in M^n_{n-1}$. 
%%% 
\end{proof}

\begin{Remark}
 Thus,~$1$-positive systems are cooperative systems, 
and~$(n-1)$-positive systems are competitive systems, so the notion of a~$k$-positive system provides a generalization of both cooperative and competitive systems. 
\end{Remark}

%%%%%%%%%%%%%%%%%%%%%%%%%%%

We now turn to consider~$A^{[k]}$ with~$k\not = n-1 $.
The case~$k=n$ is trivial as~$A^{[n]}$ is a scalar, so the associated linear dynamical system is always cooperative. The case~$k=1$ is also clear as~$A^{[1]}=A$. 
%%%
Thus, we only need to consider the case~$k\in[2,n-2]$.

\subsection{$k$-positive systems for some~$k\in[2,n-2]$}

We begin by defining a special set of periodic Jacobi matrices.
%%%%%%
\begin{Definition}\label{def:Wkset}
	%%%%% 
	For any~$k\in[2,n-2]$
	let~$M^n_k $ 
	denote the set of matrices~$A\in\R^{n\times n}$ satisfying:
	\begin{enumerate}[(a)]
		\item $(-1)^{k-1}a_{1n}, (-1)^{k-1}a_{n1} \geq 0$; 
		\item $a_{ij}\geq 0$ for all~$i,j$ with~$|i-j|=1$;
		\item $a_{ij} =  0$ for all~$i,j$ with~$1<|i-j|<n-1$.
	\end{enumerate}
\end{Definition}
%%%%%%%%%%%%%%%%%%%
For example,  
the matrices in~$M^5_3$ are those with the sign pattern:
\[
\begin{bmatrix}
*       & \geq 0 & 0 & 0& \geq 0 \\
\geq 0 & * & \geq 0 &  0&  0 \\
0& \geq 0 & * & \geq 0 &  0 \\
0& 0& \geq 0 & * & \geq 0  \\
\geq 0 & 0&  0 & \geq 0 & *  
\end{bmatrix},
\]
where~$*$ denotes ``don't care''. 
Note that the definition of~$M^n_k$ implies that~$M^n_i=M^n_j$ 
for any~$i,j \in[2,n-2]$ that have the same parity.

The next result  generalizes
 Lemma~\ref{lem:a[22]}.
%%%%
\begin{Theorem}\label{thm:explicit_a}
	%%%
	Let~$A \in \R^{n\times n  }$ with~$n>2$. 
	Then for any~$k\in[2,n-2]$
	 the matrix~$A^{[k]}$  is Metzler iff~$A\in M^n_k$.
	%%%%%%
\end{Theorem}

%%%%%%%%%%%%%%%%%%%%%%%%%%%%%%%%%%%%%%%%%%%%%
{\sl Proof of Thm.~\ref{thm:explicit_a}}.
%%%%%%%%%%%%%%%%%%%%%%%%%%%%%%%%%%%%%%%%%%%%%
We already proved this result for~$k=2$. 
Fix~$k\in[3,n-2]$. 
It follows from Lemma~\ref{lem:poltr} 
that an off-diagonal entry of~$A^{[k]}$ 
corresponding to~$(\alpha|\beta)=(i_1,\dots,i_k|j_1,\dots,j_k) $
can be nonzero  only if  all the indices in~$\alpha$ and~$\beta$ coincide, except for a single index~$i_\ell\not = j_m $, and then this entry is
$A^{[k]}(\alpha,\beta)=(-1)^{\ell+m} a_{i_\ell j_m}$.
We use this to determine when~$a_{ij}$ (or~$-a_{ij}$) appears 
on an off-diagonal entry of~$A^{[k]}$.  We   consider only
pairs~$(i,j)$ with~$j\geq i$, as the case~$i\geq j$ follows by symmetry. 

\noindent {\sl Case 1.} 
%%%%%%%%%%%%%%%
If~$ j=i$ then~$a_{ij}=a_{ii}$ does not appear in 
any off-diagonal entry of~$A^{[k]}$. This explains the ``don't care''s in the definition of~$M^n_k$.

\noindent {\sl Case 2.} 
%%%%%%%%%%%%%%%
If~$j=i+1$ then~$a_{ij}=a_{i,i+1} $ will appear 
in 
an off-diagonal entry~$(\alpha|\beta)$ of~$A^{[k]}$ if all the entries of~$\alpha$ and~$\beta$ coincide
 except that~$i$   appears in 
$\alpha$ but not in~$\beta$, and~$i+1$ appears in 
$\beta$ but not in~$\alpha$. 
But this implies that~$i$ and~$i+1$
appear in the same entry of~$\alpha$ and~$\beta$, that is,~$\ell=m$ 
and the off-diagonal entry of~$A^{[k]}$ is~$(-1)^{2 \ell}a_{i,i+1}=a_{i,i+1}$.
Hence,~$A^{[k]}$ is not Metzler if~$a_{i,i+1}<0$. 

\noindent {\sl Case 3.} 
%%%%%%%%%%%%%%%
Suppose that~$1<j-i<n-1$ and~$j=i+2$ (so~$i+2\leq n$).
 We now show that both~$a_{i,i+2}$
and~$-a_{i,i+2}$ appear on off-diagonal entries of~$A^{[k]}$. 
It is not difficult to show that since~$k+2\leq n$ and~$i+2\leq n$, there 
exists  an integer~$x$ such
 that
\begin{align}\label{eq:socupxx}
1\leq x \leq i \text{ and } i-k+1\leq x \leq n-k-1.
\end{align}
Then for
\begin{align*}
\alpha:=\{&x,\dots, i-1,i,  \widehat{i+1},\widehat{i+2},   i+3,\dots,x+k+1 \},\\
\beta:=\{ &x,\dots,i-1,     \widehat i, \widehat {i+1}, i+2,i+3,\dots,x+k+1\},
\end{align*}
where~$\widehat j$ means that~$j$ is \emph{not}
 included in the set,
we have~$A^{[k]}(\alpha |  \beta )
=a_{i,i+2}$, so~$a_{i,i+2}$   appears on an off-diagonal
 entry of~$A^{[k]}$. Note that~\eqref{eq:socupxx} guarantees that~$\alpha$ [$\beta$] includes~$i$
$[i+2]$.
 
Similarly, it 
 is not difficult to show that since~$2\leq k\leq n-1$
 and~$i+2\leq n$, there 
exists  an integer~$x$ such
 that
\begin{align}\label{eq:xcon}
1\leq x \leq i \text{ and } i-k+2\leq x \leq n-k.
\end{align}
Then for
\begin{align*}
\alpha:=\{&x,\dots, i-1,i,i+1,\widehat{i+2},  i+3,i+4,\dots,x+k \},\\
\beta:=\{ &x,\dots,i-1, \widehat{i}, i+1,i+2,i+3,\dots,x+k\},
\end{align*}
we have~$A^{[k]}(\alpha |  \beta )
=-a_{i,i+2}$, so~$-a_{i,i+2}$  also  appears on an off-diagonal
 entry of~$A^{[k]}$. 
Hence,~$A^{[k]}$ is not Metzler if~$a_{i,i+2}\not = 0$. Note that~\eqref{eq:xcon} 
guarantees that~$\alpha$ [$\beta$] 
 includes~$i$ [$i+2$].

\noindent {\sl Case 4.} 
%%%%%%%%%%%%%%%
Suppose that~$1<j-i<n-1$ and~$j>i+2$.
Then it can be shown as in Case~3   that
  both~$a_{i,j}$
and~$-a_{i,j}$ appear on off-diagonal entries of~$A^{[k]}$. 
Hence,~$A^{[k]}$ is not Metzler if~$a_{i j}\not = 0$.

\noindent {\sl Case 5.} 
%%%%%%%%%%%%%%%
Suppose that~$ j-i=n-1$, that is,~$i=1$ and~$j=n$.
Then~$a_{ij}=a_{1n}$   appears in an entry~$(\alpha|\beta)$
 of~$A^{[k]}$
only when~$\alpha=\{	1,i_2,\dots,i_k\}$ 
and~$\beta=\{	j_1,\dots, j_{k-1},n\}$,
with~$i_{p+1}=j_p$ for all~$p\in[1,k-1]$,
and then 
\[
A^{[k]}(\alpha|\beta)=(-1)^{1+k } a_{1n}.
\]
Hence,~$A^{[k]}$ is not Metzler if~$(-1)^{k-1 } a_{1n}<0$. 

Summarizing the cases above, we conclude that if~$A\not \in M^n_k$ then~$A^{[k]}$ is not Metzler. But the analysis above
 actually covers all the cases where an entry~$a_{ij}$ appears as an off-diagonal entry of~$A^{[k]}$, and this completes the proof of Thm.~\ref{thm:explicit_a}.~\hfill{$\square$}

Combining Thm.~\ref{thm:kiff} and Thm.~\ref{thm:explicit_a} yields the following result.
\begin{Corollary}
%%%%%%%%%%%%%%%%%%%%%%
For any~$k\in[2,n-1]$
the   LTV~\eqref{eq:linsysp}  is~$k$-positive 
on~$(a,b)$
iff~$A(s) \in M^n_k$ for almost
all~$s\in(a,b)$. 
%%%%
\end{Corollary}

Using the explicit structure of a~$k$-positive system yields a  generalization
of Corollary~\ref{cor:1_and_2_implies_all_k}.
%%%%%%%%%%%%%%%%%
\begin{Corollary}\label{coro:even_and_odd_implies_all_k}
Suppose that there exist~$i,j\in[1,\dots,n-2]$,
with~$i$ even  and~$ j $ odd 
   such that 
 the system~\eqref{eq:linsysp} is~$i$-positive and~$j$-positive.
Then~\eqref{eq:linsysp}  
 is~$k$-positive for all~$k\in[1,n]$.	
\end{Corollary}
%%%

\begin{proof}
Since the system is~$i$-positive with~$i$ even,   
Definition~\ref{def:Wkset} implies that:~$a_{1n},a_{n1}\leq 0$, the super- and sub-diagonals of~$A$ include non-negative entries, and 
all other off-diagonal entries are zero. The system is also~$j$-positive with~$j$ odd. If~$j=1$ then~$A$ is Metzler, so we conclude that~$a_{1n}=a_{n1}=0$.
If~$j>1$ then    
Definition~\ref{def:Wkset} implies 
 that~$a_{1n},a_{n1} \geq 0$, so again~$a_{1n}=a_{n1}=0$. We conclude that~$A$ is tridiagonal and Metzler, and 
thus~\eqref{eq:linsysp}  
 is~$k$-positive for all~$k\in[1,n]$.	
%%%%
\end{proof}

%%%%%%%%%%%%%%%%%%%%%%%%%%%%%%%%%%%%%%%%%%%%%%%%%%%%%%%%%%%%%%
\section{Geometrical structure of the invariant sets}\label{sec:geom}
%%%%%%%%%%%%%%%%%%%%%%%%%%%%%%%%%%%%%%%%%%%%%%%%%%%%%%%%%%%%%%
A natural question is what is the structure of the invariant sets~$P^k_-$ and~$P^k_+$ defined above. 
It is clear that these sets are cones, as~$s^-(x)=s^-(\alpha x)$  for all~$\alpha \in \R$,
and~$s^+(x)=s^+(\alpha x)$  for all~$\alpha \in \R\setminus\{0\}$.
%%%%%%%%%%%%%%%%%%%%%%
 However,     these sets are \emph{not} convex cones. 
For example, for~$n=2$ the vectors~$ x=\begin{bmatrix} 1& 1 \end{bmatrix}' ,y=\begin{bmatrix}  -1&-1\end{bmatrix}' $ satisfy~$x,y \in P^1_+$,
 yet~$\frac{1}{2}(x+y)=\begin{bmatrix} 0&0  \end{bmatrix}' \not \in  P^1_+ $.
Similarly, for~$n=3$ the
 vectors~$ x=\begin{bmatrix} 1&-1&0\end{bmatrix}' ,y=\begin{bmatrix} 0&-1&1\end{bmatrix}' $ satisfy~$x,y \in P^2_-$,
 yet~$\frac{1}{2}(x+y)=\begin{bmatrix} 1/2&-1&1/2 \end{bmatrix}' \not \in  P^2_- $.

Recall that a dynamical system is called \emph{monotone}
 if its flow is order-preserving 
with respect to the  (partial) order~$\leq$ induced by a closed, convex and pointed cone~$K$, that is,
\[
x\leq y \iff  y-x  \in K.
\]
The convexity of~$K$ implies that
\[
 x\leq y, \; y\leq z \implies x\leq z,
\] 
%%%
and the fact that~$K$ is pointed yields 
\[
 x\leq y, \; y\leq x \implies x=y.
\]
 Since~$P^k_-,P^k_+$ are not convex, this suggests that $k$-positive systems are not monotone. Fortunately, these sets, although not convex, do possess 
a  useful structure.

\subsection{$P^k_-$ is a  cone  of rank~$k$}
%%%%%%%%%%%%%%%%%%%%%%%%%%%%%%%%%%%%%%%

Recall that a set~$C\subseteq  \R^n$  is called a  \emph{cone of rank~$ k$} 
(see e.g.~\citep{pls_sobolev,sanchez2009cones}) 
if: \begin{enumerate}[(1)]
\item $C$ is closed, 
\item $x\in C$ implies that~$\alpha x \in C$ for all~$\alpha \in \R$,
and 
\item $C$ contains a linear subspace
of dimension~$k$ and no linear subspace of higher dimension.
\end{enumerate}
%%%%%%%%%%% 
For example, it is straightforward to see that~$\R^2_{+} \cup (-\R^2_{+})$ 
(and, more generally,~$\R^n_{+} \cup (-\R^n_{+})$)
  is  a cone of rank~$1$. 
 
A cone~$C$ of rank~$k$ is called \emph{solid} if its interior is nonempty, and
\emph{$k$-solid} if there is a linear subspace~$W$ of dimension~$k$ such that~$W\setminus\{0\} \subseteq \Int(C)$. In the context of dynamical systems,
such  cones are important because  trajectories of dynamical systems
that are confined to~$C$   can be projected to the linear subspace~$W$~\citep{sanchez2009cones}. Roughly speaking, if this 
 projection is one-to-one then the  
trajectories must satisfy the same properties as trajectories in a $k$-dimensional space. 
 \cite[Ch.~1]{pls_sobolev} showed
that the set~$P^k_-$   
is a~$k$-solid   cone. The next result slightly strengthens this. 
Also, the proof, unlike that in~\citep{pls_sobolev},
uses the elegant spectral properties of   oscillatory matrics. 
 
 %%%%%%%%%
\begin{Lemma}
%%%%%%%%%%%%%%%%%%%%%%%%%%%%%%%%%%%%%%%%%%%%
For any~$k\in[1,n-1]$ the set~$P^k_-$   
is a~$k$-solid   cone, and its complement 
\be\label{eq:defpkmc}
( P^k_-)^c:= \Clos(\R^n\setminus P^k_-) 
\ee
is an~$(n-k)$-solid cone.
%%%%%%%%%%%%%%%%%%%%%%%%%%% 
\end{Lemma}
%%%%%%%%%%
{\sl Proof.} 
%%%%%%%%%%%%%%%%%%%%%%%%%%
Pick~$k\in[1,n-1]$. It follows from the definition of~$s^-$ that~$P^k_-$ is closed.
If~$x\in P^k_-$, that is,
$s^-(x)\leq k-1 $ then clearly~$\alpha x \in  P^k_-$ for all~$\alpha \in \R$. 
The set~$P^k_-$ cannot contain a linear subspace of dimension~$k+1$, as
using a linear combination of~$k+1$ independent vectors in~$\R^n$  one can generate a vector~$y$ such that~$s^-(y)\geq k$. 
Let~$A\in\R^{n\times n}$ be an oscillatory matrix, and denote its eigenvalues and eigenvectors as in Thm.~\ref{thm:spec}. Then~\eqref{eq:svano} implies that
 for any~$c_1,\dots c_k\in\R$, that are not all zero,
%%%%%%%%%%%%%
\be\label{eq:bnpip}
s^-(\sum_{p=1}^k c_p u^p)\leq s^+(\sum_{p=1}^k c_p u^p) \leq k-1.
\ee
We conclude that~$W:=\text{span}\{u^1,\dots,u^k\}\subseteq P^k_-$, and that~$W\setminus\{0\}\subseteq  P^k_+$.

Now pick~$x\in W  $. Suppose  that~$x\in\partial P^k_-$.
Then by the definition of~$s^-$,~$x$ includes a zero entry, say,~$x_i$ and there exists~$\varepsilon \in \R  \setminus\{0\}$,
with~$|\varepsilon|$ arbitrarily small, 
such that the vector~$\tilde x$ obtained from~$x$ by setting~$x_i$ to~$\varepsilon$
 satisfies~$s^-(\tilde x) > k-1$. Thus,~$s^+(  x)>k-1$. But now~\eqref{eq:bnpip}
	gives~$  x =0$. We conclude that~$ W \cap  \partial P^k_- =\{0\}$.
	This   shows that~$ W\setminus\{0\} \subseteq \Int(P^k_-)$, so~$P^k_-$ is 
  a~$k$-solid   cone.

	We now turn to prove the assertion for~$(P^k_-)^c$. By definition, 
	this set is closed, and~$x\in (P^k_-)^c$ implies that~$\alpha x \in (P^k_-)^c$ for all~$\alpha\in \R$.  
	 %%%%%%
	Eq.~\eqref{eq:svano} implies that for any~$c_{k+1},\dots c_n\in\R$, that are not all zero,
\be\label{eq:gysp}
     k\leq s^-(\sum_{p=k+1}^n c_p u^p)  .
\ee
In other words, for~$  W^c :=\text{span} \{u^{k+1},\dots,u^n \}$ 
we have~$ W^c \setminus\{0\}
\subseteq \R^n\setminus P^k_-$. Combining this with~\eqref{eq:defpkmc}
	implies that~$W^c \subseteq ( P^k_-)^c$.
	
	Pick~$x\in   W^c$, that is, $x=\sum_{p=k+1}^n d_p u^p$, for some~$d_{k+1},\dots,d_n \in \R$.  
	Suppose that~$x\in \partial( ( P^k_-)^c)$. Since~$P^k_-$ is closed, we conclude that~$x\in \partial  P^k_-$. 
	Thus,~$x\in \{0\} \cup \{x\in \R^n: s^-(x)=k-1\} $. 
If~$x\not =0$ then at least one of the~$d_i$s is not zero, so~\eqref{eq:gysp}  yields
\[
      k \leq s^-(x) =k-1.
\]
We conclude that~$x=0$, so~$W^c \setminus \{0\}
	\in \Int(( P^k_-)^c)$. 
%%%%%%%%%
Thus,~$ ( P^k_-)^c$
	is an~$(n-k)$-solid cone.~\hfill{$\square$}
%%%%%%%%%%%%%%%%%%%%%%%%%%% 

	Our next goal is to derive an explicit decomposition
	for the sets~$P^k_-$, $ P^k_+$. 
%%%%%%%%%%%%%%%%%%%%%%%%%%%%%%%%%%%%%%%%%%%%%%%%%%%
\subsection{$P^k_-$  is the union of convex sets}
%%%%%%%%%%%%%%%%%%%%%%%%%%%%%%%%%%%%%%%%%%%%%%%%%%
For any~$k\in[1,n]$, define
\[
Q^k_-:=\{ z\in\R^n : s^-(z) = k-1 \} .
\]
For example~$Q^1_- =\R^n_+ \cup  \R^n_-$, and~$Q^2_-= F \cup (-F)$,
where~$F$ is the set of all
 vectors with the sign pattern
\[
\geq 0,\dots,\geq 0,\leq 0,\dots,\leq 0,
\]
with  at least one entry   positive and one entry   negative. 
Note that~$x \in Q^k_-$ implies that~$\alpha x \in  Q^k_-$ for all~$\alpha\in\R\setminus\{0\}$.

Any vector~$y\in Q^k_-$ can be decomposed into~$k$ disjoint and consecutive 
sets of entries, where 
each set is composed of entries that are all nonnegative [nonpositive]
and at least one entry is positive  [negative].
 For example, the vector~$y=\begin{bmatrix} 
0&1&2&0&-2& 0&1 &2
\end{bmatrix}'$ satisfies~$y\in Q^3_-$ and can be decomposed into three sets:
the first is
$0,1,2,0$, the second is~$-2, 0$, and the third is~$1, 2$. 
We use this idea  to derive a decomposition of~$Q^k_-$. We require the following definition. 
\begin{Definition}
For a vector~$v=\begin{bmatrix} v_1&\dots&v_k\end{bmatrix}' $ with integer entries such that
\be\label{eq:vvec}
1\leq v_1<v_2<\dots<v_k=n,
\ee
  let~$C^k_-(v)  \subseteq \R^n$ denote   the set 
	of all   vectors~$y\in\R^n $ satisfying: 
\begin{itemize}
\item
 $y_1,\dots,y_{v_1}\geq 0$, with at least one of these entries positive;
%%%
\item $y_{v_1+1}<0$, and~$y_{v_1+2},\dots, y_{v_2}\leq 0$;
\item $y_{v_2+1}>0$, and~$y_{v_2+2},\dots, y_{v_3}\geq 0$; and so on until
%%%%%%%%%%%%%%%%%%%%%%%%%%%
\item $(-1)^{k-1} y_{v_{k-1}+1} >0 $,
and $(-1)^{k-1} y_{v_{k-1}+1}$,$\dots$, $(-1)^{k-1} y_{v_{k}} \geq 0 $ 
(recall that~$v_k=n$).  
\end{itemize}
\end{Definition}
 
%%%%%%%%%%%%%%%%%%%%%%%5
For example, for~$n=4$, $k=3$, and~$v=\begin{bmatrix}2& 3&  4\end{bmatrix}'$, 
\[
C^3_-(v)= \{y\in \R^4: y_1\geq 0, y_2 \geq 0, y_1 y_2\not =0, y_3<0,y_4>0 \}.
\]
Note that~$C^k_-(v)$ is a convex cone. In fact,~$C^k_-(v)$ is an orthant in~$\R^n$,
and if~$i\not = j$ then~$C^i_-(v)$ and~$C^j_-(v)$ 
are different orthants.

It is clear that~$y\in Q^k_-$ iff~$y\in C^k_-(v) \cup (- C^k_-(v))$
 for some~$v=\begin{bmatrix}
v_1&v_2&\dots&v_k\end{bmatrix}'$
 satisfying~\eqref{eq:vvec}. The number of different vectors~$v$ that
 satisfy~\eqref{eq:vvec} is~$\binom{n-1}{k-1}$, as we fix~$v_k=n$.
Combining this with the
definitions of~$P^k_-$ and~$Q^k_-$ yields
the following 
  characterization of~$P^k_-$ as the union of convex 
  cones.

\begin{Proposition} \label{prop:geometrical_sturdcture}
%%%%%%%%%%%%%%%%%%%%%%%%%%%%%%%%%%%%%%%%%%%%%%%%%%%%%%%%%%%%%%%%
For any~$s\in[1,n]$ we have 
	\[
	%%%%
								P^s_- = \bigcup_{k=1} ^ s Q^k_-,
	\]
where 
 \[
	%%%%
								Q^k_- = \bigcup_{i=1} ^ {\binom{n-1}{k-1}} C^k_-(v^i) \cup ( -C^k_-(v^i)),
	\]
	and~$v^i$, $i\in [1,\binom{n-1}{k-1}]$,
	are all the different  vectors
	that satisfy~\eqref{eq:vvec}.
	 %%%%%%%%%%%
\end{Proposition}
%%%%%%%%%%%%%%%%%%%%%%%%%%%%%%%%%%%%%%%%%%%%%%%%%%%%%%%%%

 \begin{Example}\label{exa:sminuin_2}
%%%
Consider again the trajectory~$x(t)$
 of the system in
Example~\ref{exa:sminuin} with
\[
x(0)=\begin{bmatrix}   
0.34 &   -0.54&   -1.06&    0.49   \end{bmatrix}'.
\]
%%%  
%%%
Recall that here~$s^-(x(t))\leq 2 $ for all~$t\geq 0$.
Note that~$x(0)\in C^3_-  (\begin{bmatrix}  1&3&4 \end{bmatrix}')$. 
An analysis of this trajectory shows that it crosses through 
 the following cones:
\begin{align*}
C^3_- &(\begin{bmatrix}  1&3&4 \end{bmatrix}')
\to
C^3_-(\begin{bmatrix}  2&3&4 \end{bmatrix}')
\to 
C^1_-(\begin{bmatrix}   4 \end{bmatrix}  )\\&
\to
C^3_-(\begin{bmatrix}  1&2&4 \end{bmatrix}')
\to
C^2_-(\begin{bmatrix}  2&4 \end{bmatrix}').
%%%%%
%%%%%%%%%%%%
\end{align*}
Note that all these cones belong to~$P^3_-$. 
\end{Example}

%Recall that~$C^k_-(v)$ is an orthant in~$\R^n$,
%and that~$C^i_-(v)$, $C^j_-(v)$, with~$i \neq j$,
%are different orthants.
%It follows that~$D(C^i_-(v))$ and~$D(C^j_-(v))$, with~$i \neq j$,
%are different orthants in~$\R^n$.

\begin{Remark}
%%%%%%%%%%%%%%%%%
The duality relation~\eqref{eq:s_minus_s_plus} and the fact that~$D^{-1}=D$
implies that
\begin{align}\label{eq:duliop}
D P^k_-&:= \{D x: x\in\R^n,\;  s^-(x)\leq k-1\} \nonumber \\
     &=\{x\in\R^n: s^-(Dx)\leq k-1\}\nonumber\\
     &=\{x\in\R^n: s^+( x)\geq n-k\}\nonumber\\
		     &=\R^n\setminus \{x\in\R^n: s^+( x) <  n-k\}\nonumber\\
		     &=\R^n\setminus \{x\in\R^n: s^+( x) \leq   n-k-1\}\nonumber\\
				&=\R^n\setminus P^{n-k}_+ .
			%%	&=\R^n\setminus \Int (P^{n-k}_-).
%%%%%
\end{align}
Thus, the  results above on the structure of~$P^k_-$,~$k\in[1,n-1]$ 
can be transformed to characterizations of~$P^{j}_+$, 
$j\in \{n-1,n-2,\dots, 1\}$,  using~\eqref{eq:duliop}.
%%%%%%%%%%%%%
For example,
%%%%
since~$P^1_-=\R^n_+ \cup \R^n_-$, \eqref{eq:duliop} implies that
\[
P^{n-1}_+=\R^n\setminus ((D\R^n_+) \cup (D\R^n_-)) . 
\]
In other words,~$P^{n-1}_+$ 
is the set of all vectors except for those with either
the sign pattern~$\begin{bmatrix} \geq 0& \leq 0& \geq 0 &\dots
\end{bmatrix}'$ or the sign pattern~$\begin{bmatrix} \leq 0& \geq 0& \leq 0 &\dots
\end{bmatrix}'$.
%%%
%%%
%%%%
\end{Remark}

Note that~\eqref{eq:duliop} implies that in general the sets~$P^k_-$ and~$P^j_+$ have a different structure.
For example,~$P^k_-$ is closed for every~$k$ so~\eqref{eq:duliop} implies that~$P^j_+$ is open for every~$j$. 
Also,~$0\in P^k_-$ for all~$k\in[1,n]$, so~$0\not \in P^k_+$ for all~$k \in[1,n-1]$.

The next section
  describes several applications of the notion of~$k$-positive linear systems to the asymptotic
	analysis of nonlinear dynamical 
	systems.
%%%%%%%%%%%%%%%%%%%%%%%%%%%%%%%%%%%%%%%%%%%%%%%%%%%
\section{Applications  to nonlinear dynamical systems}\label{sec:applnon}
%%%%%%%%%%%%%%%%%%%%%%%%%%%%%%%%%%%%%%%%%%%%%%%%%%%
We begin by considering    time-varying 
  nonlinear systems, and then results for
the time-invariant   case  follow as a special case. 

Consider the   time-varying  nonlinear dynamical system:
%%%%%%%%%%%%%%%%%%%%%%%%%%%%%%
\begin{equation} \label{eq:nonlinear}
    \dot x(t)=f(t,x(t)),
\end{equation} 
%%%%%%%%%%%%%%%%%%%%%%%%%%%%%%%%%%%%
whose trajectories  evolve  on a convex
 invariant 
set~$\Omega \subseteq \R^n$. 

We assume throughout that~$f$ is~$C^1$ with respect to its second variable~$x$,
and that for all~$z\in\Omega$ the map~$t\to f(t,z)$ is measurable and essentially bounded. 
Denote the Jacobian of~$f$ with respect to its second variable
by~$J(t,x):=\frac{\partial}{\partial  x} f(t,x)$. 

For any initial condition~$x_0\in\Omega$ and any initial time~$t_0 \in(a,b)$   
we assume throughout that~\eqref{eq:nonlinear}
  admits
a unique solution for all~$t\geq t_0$ 
and denote this solution 
 by~$x(t,t_0,x_0)$. In what follows we take~$t_0=0$ and write~$x(t,x_0)$ 
for~$x(t,0,x_0)$. 
%%%%%%%%%% 

The application of~$k$-positive linear
 systems to~\eqref{eq:nonlinear}
 is based on the variational system associated with~\eqref{eq:nonlinear}. To define this, 
fix~$p,q\in \Omega$.
Let~$z(t):=x(t,p)-x(t,q)$, and for~$r\in[0,1]$, 
 let~$\gamma(r):=r x(t,p)+(1-r)x(t,q)$.
 Then
\begin{align*} 
%%%%%%%%%%%
   \dot z(t)&=f(t,x(t,p))-f(t,x(t,q))  \\
	       &=\int_0^1  \frac{\partial }{\partial r}   f(t,\gamma(r))     \dif r  ,
				%% &=\int_0^1  J(t,\gamma(r))       ( x(t)-y(t)   ) \dif r \\
	%%%%%				&=A(t)z(t),
%%%%%
\end{align*}
%%%
and this gives the LTV:
\be\label{eq:ltvva}
\dot z(t)=A^{pq}(t)z(t),
\ee
with
\begin{equation}\label{eq:at_int}
    A^{pq}(t):=\int_0^1  J(t,\gamma(r))     \dif r.  
\end{equation}
%%%
%%
 This LTV is the \emph{variational system}
 associated with~\eqref{eq:nonlinear}. 
\begin{Definition}
We say that the nonlinear system~\eqref{eq:nonlinear} is \emph{[strongly] $k$-cooperative} 
if the LTV~\eqref{eq:ltvva} is [strongly]~$k$-positive for all~$p,q\in\Omega$.
%%%%%
\end{Definition}

The results above can be used to provide simple to verify sufficient conditions for
[strongly] $k$-cooperativity  
 of~\eqref{eq:nonlinear}. The next two results demonstrate  this. 
\begin{Corollary}
%%%%%%%%%%%%%%%%%%%%%
Suppose that there exists~$k\in[1,n-1]$ such 
that~$J(t,z)\in M^n_k$ for almost all~$t\in(a,b)$ and all~$z\in\Omega$. 
Then~\eqref{eq:nonlinear} is $k$-cooperative  on~$(a,b)$.
If, furthermore,   for any~$z\in \Omega$ and any~$a<t_0<t<b$
there exists~$\tau\in[t_0,t)$ such that~$J(t,z)$ 
is irreducible at~$\tau^+$ 
then~\eqref{eq:nonlinear} is strongly $k$-cooperative 
 on~$(a,b)$.
%%%%%%%%%%%%%%%%%%%%%%%
\end{Corollary}
The proof follows from the fact that, by the definition of~$M^n_k$, if~$F,G\in M^n_k$ 
then~$F+G\in M^n_k $, and this is carried over to the integration in~\eqref{eq:at_int}.  
Also, addition of  two matrices in~$M^n_k$   cannot change a nonzero entry
 to a zero entry, and this implies that
irreducibility is also carried over to the integral.

The next two examples describe specific examples 
 of nonlinear systems that are~$k$-cooperative 
 for some~$k$.

\begin{Example}
 \cite{Elkhader1992}
studied the   nonlinear system
\begin{align}\label{eq:alexsys}
%%%%%
\dot x_1&=f_1(x_1,x_n),\nonumber\\
\dot x_i &= f_i(x_{i-1},x_i,x_{i+1}),
\quad i=2,\dots,n-1,\nonumber\\
\dot x_n&=f_n(x_{n-1},x_n). 
%%%%%
\end{align}
It is assumed that the state-space~$\Omega\subseteq\R^n$
 is convex, that~$f_i\in C^{n-1}$, $i=1,\dots,n$, and that
there exist~$\delta_i\in\{-1,1\}$, $i=1,\dots,n$,
 such that
\begin{align*}
%%%%%
\delta_1\frac{\partial }{\partial x_n}f_1(x)   &>0,\\
\delta_2\frac{\partial }{\partial x_1}f_2(x) , \delta_3\frac{\partial }{\partial x_3}f_2(x)&>0,\\
&\vdots\\
\delta_{n-1} \frac{\partial }{\partial x_{n-2}}f_{n-1}(x),
\delta_n \frac{\partial }{\partial x_{n}}f_{n-1}(x)
 &>0,\\
\delta_n \frac{\partial }{\partial x_{n-1}}f_n(x)&>0,
%%%%
%%%%%
\end{align*}
for all~$x\in\Omega$. 
This is a generalization of     the
monotone cyclic feedback system analyzed  in the seminal work of~\cite{poin_cyclic}.
As noted by~\cite{Elkhader1992}, we may assume without loss of generality that~$\delta_2=\dots=\delta_n=1$ and~$\delta_1 \in \{-1,1\}$. Then the Jacobian of~\eqref{eq:alexsys} has the form
\[
J(x)=\begin{bmatrix}
*& 0 &0 &0 &\dots & 0 & 0  & \sgn(\delta_1) \\
%%%%%%%%%%%%%%%%%%%%%%%%%%%%%%%%%%%%%%%%%
>0 & * &>0 &0 &\dots & 0& 0 & 0  \\
%%%%
0& >0 & * &>0  &\dots &0&  0 & 0 \\
%%%
&&&\vdots\\
%%%%%%%%%%%%%%%%%%%%%%%%%%%%%
0&  0 & 0 & 0  &\dots & 0  &>0&  * \\
\end{bmatrix} ,
\]
for all~$x\in \Omega$. Note that~$J(x)$ is irreducible for all~$x\in \Omega$. 
If~$\delta_1=1$ then~$J(x)$ is Metzler, so the system is strongly~$1$-cooperative. Consider the case~$\delta_1=-1$.
Then~$J(x) \in M^n_{2}$, so the system is strongly $2$-cooperative.
(If~$n$ is odd then~$J(x) \in M^n_{n-1}$, so the system is also strongly competitive.) The main result in~\cite{Elkhader1992}
is that when~$\delta_1=-1$ the omega-limit set of any bounded solution of~\eqref{eq:alexsys}
   includes at least one   equilibrium or a 
periodic orbit. Our main result in this section generalizes this in several ways: first, 
we allow~$f_1$ [$f_n$] to depend also on~$x_2$
[$x_{n-2}$]. Second, we require~$f_i\in C^1$ for all~$i$
 rather than~$f_i\in C^{n-1}$ for all~$i$, and third we require~$J(x)$  to be irreducible, but not necessarily of the form 
assumed by~\cite{Elkhader1992}.
%%%%%%%%%%
%%%%%%%%%
\end{Example}

\begin{Example}
Our second example is a system  with scalar nonlinearities:
%%%%%%%
 %%%%%%%
%%
\be\label{eq:dofty}
\dot x(t) =C(t) \begin{bmatrix} f_1(x_1(t))  
\\f_2(x_2(t))\\\vdots\\f_n (x_n(t))\end{bmatrix},
\ee
where~$f_i:\R\to\R$,~$i\in[1, n]$,
 are~$C^1$ functions,
and~$C:(a,b)\to\R^{n\times n }$. Suppose that its trajectories evolve on a compact and convex state-space~$\Omega$. 
%%%%%		
The Jacobian of~\eqref{eq:dofty}  is
\be\label{eq:jcsdop}
J(t,x)= C (t)  \diag(f'_1(x_1)),\dots,f'_n(x_n))   , 
%%%
\ee
where~$ f'_i(z):=\frac{d}{d z} f_i(z)$. 
Pick~$p,q\in\Omega$ and consider 
 the line~$\gamma(r):=r p +(1-r)q$, $r\in[0,1]$. 
Substituting~\eqref{eq:jcsdop}  in~\eqref{eq:at_int}
 yields
%%%%%%%%%
\begin{align}\label{eq:jacoiop}
                A^{pq}(t)
                &= C (t) \diag(g_1(p_1,q_1) ,\dots,g_n(p_n,q_n) ),
\end{align}
   where
\begin{align*}
 g_i(p_i,q_i) 
&:= \begin{cases}   \frac{f_i(p_i)-f_i(q_i)}{p_i -q_i}, & \text{ if }  
p_i\not= q_i,\\
       f'_i(q_i),  & \text{ if } p_i = q_i.                           
\end{cases}
			%%%%%%%
\end{align*}
%%%%%%%
This implies that for any~$k\in[1,n-1]$ 
it is straightforward  to provide sufficient  conditions guaranteeing
that~$A^{[k]}(t)$ is Metzler. 
 To demonstrate this, 
  assume for simplicity  that 
 \[
f'_i(z) >0 \text{ for all }z\in\R \text{ and all } i\in[1,n]. 
\]
Then the compactness of~$\Omega$
implies that
 there exists~$\delta>0$ such that~$g_i(p_i,q_i)\geq \delta$ for all~$p,q\in \Omega$ and
 all~$i\in[1,n]$.
Now~\eqref{eq:jacoiop}
implies that every entry of~$A^{pq}(t)$
satisfies~$a_{ij}(t)=c_{ij}(t) m(t)$ with~$m(t)\geq \delta$ for all~$t$. 
Thus, if~$C(t)\in M^n_k$ for almost all~$t$
then so does~$A^{pq}(t)$, and~\eqref{eq:dofty} is~$k$-cooperative.  
%%%%%%%%%%%%%%%
\end{Example}
%%%%%%%%%%%%%%%%
 
We now describe several applications of  $k$-cooperativity
of~\eqref{eq:nonlinear}.
The first     is the existence of certain \emph{explicit}  invariant sets. 
The second application is less immediate
  and concerns the Poincar{\'e}-Bendixson property
	in strongly~$2$-cooperative systems.

%%%%%%%%%%%%%%%%%%%%%%%%%%%%%%%%%
\subsection{Invariant sets}
%%%%%%%%%%%%%%%%%%%%%%%%%%%%%%%%%
\begin{Proposition} \label{prop:inhgt}
%%%%%%%%%%%%%
Suppose that~\eqref{eq:nonlinear}  
is~$k$-cooperative.
%%%%
%%%%
Then for any~$p,q\in\Omega$ we have
\be\label{eq:ploip}
p-q\in P^k_- \implies  x(t,p)-x(t,q) \in P^k_-  \text{ for all } t \geq  0. 
\ee
%%%%

If furthermore~$0 \in \Omega $ and~$0$
is an equilibrium point of~\eqref{eq:ltvva}, i.e.~$f(t,0)=0$ for all~$t$ then 
\be\label{eq:ploip2}
p \in P^k_- \implies  x(t,p)  \in P^k_-  \text{ for all } t \geq  0. 
\ee

%%%%%%%%%%%%%%%%%
\end{Proposition}
%%%%%%%%% 
\begin{proof}
The proof of~\eqref{eq:ploip} follows immediately from the fact 
that~$k$-positivity of~\eqref{eq:ltvva}
			implies that 
  for any~$z( 0) \in P^k_-$  
	we have~$z(t)\in P^k_-$ for all~$t \geq  0$.
 Taking~$q=0$ in~\eqref{eq:ploip} yields~\eqref{eq:ploip2}.
%%%%%%%%%%%%%%%%%
\end{proof}

If we strengthen the requirement  
to \emph{strongly} $k$-cooperativity 
 then 
we can strengthen~\eqref{eq:ploip} to
\[
p-q\in P^k_- \setminus \{0\}  \implies  x(t,p)-x(t,q) \in P^k_+  \text{ for all } t> 0, 
\]
and~\eqref{eq:ploip2} to
\[
p\in P^k_- \setminus \{0\}  \implies  x(t,p)  \in P^k_+  \text{ for all } t> 0.
\]

Note that Prop.~\ref{prop:geometrical_sturdcture}
  provides an \emph{explicit}
	characterization of the invariant sets
	here as the union of convex sets.

Our next goal is to combine the results
 in~\citep{sanchez2009cones,feng2017}
with the facts that~$P^2_-$ is~$2$-solid
and its complement~$(P^2_-)^c$ is~$(n-2)$-solid to establish the Poincar\'{e}-Bendixson property
for systems that are
 strongly~$2$-cooperative. 
%%%%%%%
The next remark states a key point that allows us to prove a result that is considerably stronger \updt{than} that in~\citep{sanchez2009cones}.
	\begin{Remark}\label{re:2i}
%%%%
Suppose that the nonlinear system~$\dot x=f(x)$
   is~$2$-cooperative,
i.e.~$J(x) \in M^n_2$  for all~$x\in \Omega$. It follows from the definition of the sets~$M^n_k$ that~$J(x)\in M^n_i$ for~$i=2,4,6,\dots$ and
	   all~$x\in \Omega$. Thus, the system is in fact~$(2i)$-cooperative for all~$i\geq 1$. Similarly, 
		 strongly $2$-cooperativity implies 
strongly~$(2i)$-cooperativity for all~$i\geq 1$.
%%%%%%%%%%%%%%
\end{Remark}

It is important to note that the framework of~$k$-cooperative systems cannot be used to analyze stability
and not even boundness. 
Indeed, consider  the LTI~$\dot x=Ax$.  
The conditions for~$k$-positivity  do not depend on the diagonal entries of~$A$, so the system is~$k$-positive iff~$\dot x=(cI+A)x$ is~$k$-positive for any~$c\in\R$.

 However, as we will
 see below~$2$-cooperativity   has important implications on the possible asymptotic behavior of any \emph{bounded} solution. 
For simplicity, we sometimes 
consider systems whose trajectories evolve 
 on a compact set, so  that every solution is bounded. 
Alternatively, the results hold  for  \emph{any  bounded} trajectory.

%%%%%%%%%%%%%%%%%%%%%%%%%%%%%%%%%%%%%%%%%%%%
%%%%%%%%%%%%%%%%%%%%%%%%%%%%%%%%%%%%%%%%%%%%%%%%
\subsection{Poincar\'{e}-Bendixson property}
%%%%%%%%%%%%%%%%%%%%%%%%%%%%%%%%%%%%%%%%%%%%%%%%
We begin by recalling some definitions and results by~\cite{sanchez2009cones}. 
  Let~$C\subseteq\R^n$ be a~$k$-solid cone.
	A   set~$S\subset \R^n$ is called \emph{strongly ordered} if any~$v,w\in S$, with~$v\not=w$, satisfy~$w-v\in \Int(C)$. 
	%We write~$x \sim y$ if~$x-y \in C$ 
	%and~$x \approx y$ if~$x-y \in \Int(C)$ (note that these are not   order relations, as~$C$ is not necessarily convex nor pointed).
	 A map~$M:\R^n \to \R^n$ is called \emph{positive} 
	 if~$M C \subseteq  C$,  and \emph{strongly positive} if~$M(C\setminus\{0\}) \subseteq \Int(C)$.
%%%
%%%
	Consider the time-invariant
	dynamical system~$\dot x=f(x)$ and the associated variational equation~$\dot z (t)=A^{pq}(t)z(t)$, with
$    A^{pq}(t):=\int_0^1  J(r x(t,p)+(1-r)x(t,q) )     \dif r.  
$
The nonlinear system is said to be~\emph{$C$-cooperative}
 if~$A^{pq}(t)$ is strongly positive for all~$p,q$ in the state-space
 and all~$t>0$. 
A solution~$x(t,x_0)$ is called \emph{pseudo-ordered}
if there exists a time~$\tau\geq 0 $
 such that~$\dot x(\tau,x_0)\in \Int (C)$. 
Note that since~$z(t):=\dot x(t)$ satisfies the variational equation, this implies that~$\dot x(t,x_0) \in \Int(C)$ for all~$t\geq \tau$.

	The main  result in~\citep{sanchez2009cones} 
	establishes  a strong  Poincar\'{e}-Bendixson property
	for pseudo-ordered  solutions of a C-cooperative system. 
	\begin{Theorem}~\citep{sanchez2009cones} \label{thm:sancg}
	%%%%
	 Suppose that
	the dynamical system~$\dot x=f(x)$ is~$C$-cooperative 
	with respect to a~$2$-solid cone~$C\subseteq\R^n$  whose 
  complement~$\Clos(\R^n\setminus C)$   is~$(n-2)$-solid.
Let~$x(t,x_0)$ be a  solution  with a compact omega-limit set~$\omega(x_0)$
and suppose that~$\dot x(\tau,x_0)\in C$ for some~$\tau\geq 0$.
If~$\omega(x_0)$ does not include an equilibrium 
then it is a closed orbit.
	%%%%
	\end{Theorem}
	An important tool in the
	proof of this result is~$\mathcal P:\R^n\to W$
	the linear projection  onto~$W$, parallel to the complement~$W^c$, where~$W$ is a~$2$-dimensional subspace contained in~$C$.
	\cite{sanchez2009cones}
	proved that if the pseudo-ordered solution is  a 
	closed orbit~$\gamma$ then~$\gamma$ is strongly ordered, and deduces that
	the projection~$\mathcal  P$ of~$\gamma$ is one-to one. 
	He then uses the closing lemma~\citep{Arnaud1998} 
	to extend the results to pseudo-ordered solutions that are not necessarily closed orbits. 
	%Letting~$z:=\dot x$, we have~$
	%\dot z=J(x)z$, and since~$z(\tau) \in C$, this implies that~$z(t)\in \Int( C)$ for all~$t>\tau$. This can be used to show that the projection along the
	%trajectory  is one-to-one. 

	We can now state  the main result in this section. 
	%%%%%%%%%%%%%%%%%%%
		\begin{Theorem}  \label{thm:2dim}
	%%%%
	 Suppose that
	the   system~$\dot x=f(x)$ is strongly~$2$-cooperative.
Let~$x(t,x_0)$ be a  solution  with a compact omega-limit set~$\omega(x_0)$.
If~$\omega(x_0)$ does not include an equilibrium 
then it is a closed orbit.
	%%%%
	\end{Theorem}

Note that this result is considerably
stronger than Thm.~\ref{thm:sancg}, as it applies to \emph{any} solution with a compact omega-limit set and not only to pseudo-ordered solutions. 
	Note also that the explicit analysis of the set~$M^n_2$ can be immediately 
	used to provide  a simple
	condition  for strongly~$2$-cooperativity in terms of the
	sign pattern of the Jacobian~$J(x):=\frac{\partial}{\partial x}f(x)$.
 	Note also that we have an explicit expression for a set of vectors that span a 2-dimensional subspace in~$P^2_-$ (in terms of eigenvectors of an oscillatory matrix) and thus an explicit expression for the linear projection~$\mathcal P$.
	
	The proof of Thm.~\ref{thm:2dim}
	requires several auxiliary  results.
	The next two  results analyze  
	solutions that are closed orbits.
	
	%%%%%%%%%%%%%%%%%
\begin{Lemma}\label{lemma:periodicsol2}
%%%%%%%%%%%%%%%%%%%%
 Suppose that
	the   system~$\dot x=f(x)$ is strongly~$2$-cooperative.
Let~$\gamma$ be a closed orbit corresponding to a periodic solution i.e.~$x(t+T,x_0)=x(t,x_0)$ 
for all~$t\geq 0$,
where~$T>0$ is the minimal period.
%%%%% 
Fix an even integer~$k\geq 2$. 
 If~$\dot x(\tau,x_0) \in P^k_-$ for some~$\tau\geq0 $ then
\be\label{eq:pgt2}
 x(t_2,x_0)-x(t_1,x_0) \in P^k_+    \text{ for all } 
0< t_2-t_1 <T .
\ee
%%%%%%%%%%%%%%%%%%%%%%%%%%%%%%%%%%%%
 Conversely, if~$\dot x(\tau,x_0) \not \in P^k_-$ for all~$\tau\geq0 $ then
\be\label{eq:pgt4}
%% x(t_2,x_0)-x(t_1,x_0) \not \in P^k_+  \text{ for all } 
 x(t_2,x_0)-x(t_1,x_0) \not \in P^k_-  \text{ for all } 
0< t_2-t_1 <T .
\ee
%%%%
\end{Lemma}
%%%%%%%%%%%%%%%%%%%%%%%%%%
\begin{proof}
%%%%%%%%%%%%%%%%%%%%%%%
	Since  the system is strongly~$2$-cooperative,
	it is in fact strongly~$(2i)$-cooperative for all~$i\geq 1$. 
%%%
%%%%		
Fix an even integer~$k\geq 2$. 
Suppose that there exists~$\tau \geq 0$  such that~$\dot x(\tau,x_0) \in P^k_- $. Pick~$t>\tau$. 
	Since the system is strongly~$k$-cooperative,
	$
\dot x(t,x_0) \in P^k_+, 
$
  so 
%%%
\be\label{eq:ytre}
x(t+\varepsilon,x_0) - x(t,x_0) \in P^k_+
\ee
for all~$\varepsilon>0$ sufficiently small. 
 Seeking a contradiction, assume that there 
exist two distinct points~$p,q \in \gamma$
such that
$p-q \not \in P^k_+$. 
Let~$\tau_1,\tau_2$ be such that~$0<\tau_2-\tau_1<T$,
$x(\tau_1,x_0)=q$ and~$x( \tau_2,x_0)=p$. 
Note that by adding a multiple of~$ T$ to~$\tau_1,\tau_2$ we may assume that~$\tau_1,\tau_2>\tau$. Combining this
 with~\eqref{eq:ytre} implies that
   we may actually assume that
\be\label{eq:potrew12}
					p-q \in \partial P^k_+ \subset P^k_-
\ee
and since~$P^k_+$ is an open set,
\[
p-q \not \in P^k_+.
\]
Let~$z(t):= x(t,p)-x(t,q)$. 
 Then
\[
\dot z(t)=M(t)z(t),
\]
with~$M(t):=\int_0^1 J( rx(t,p)+(1-r)x(t,q) ) \diff r $. 
Note that~$M(t)$ satisfies the same sign pattern  
as~$J$ does. Thus, if~$z(\tau)\in P^k_-$ for some~$\tau\geq 0 $ then~$z(t)\in P^k_+$ for all~$t>\tau$.
Eq.~\eqref{eq:potrew12} implies that~$z(0)\in P^k_-$, so~$z(t)\in P^k_+$ for all~$t>0$ and in particular
$
z(T) \in P^2_+
$. Thus,~$p-q \in P^k_+$. 
 This contradiction implies that
for any~$p,q \in \gamma$ with~$p\not =q$ we have
\[
 p-q \in P^k_+,
\]
and this proves~\eqref{eq:pgt2}. 

To prove~\eqref{eq:pgt4}, assume that~$\dot x(t,x_0) \not \in P^k_-$    for all~$t$, i.e.
\[
				s^-(	\dot x(t,x_0) )>k-1 \text{ for all } t. 
\]
Fix~$t\geq 0$. Then
\be\label{eq:ytre2}
 s^- ( x(t+\varepsilon,x_0) - x(t,x_0)  ) >k-1
\ee
for all~$\varepsilon>0$ sufficiently small. Thus,
 $
  x(t+\varepsilon,x_0) - x(t,x_0)   \not \in P^k_-$
for all~$\varepsilon>0$ sufficiently small.  
%%%%
 Seeking a contradiction, assume that there 
exist two distinct points~$p,q \in \gamma$
such that
\[
p-q   \in P^k_-.
\] 
Let~$\tau_1,\tau_2$ be such that~$0<\tau_2-\tau_1<T$,
$x(\tau_1,x_0)=q$ and~$x( \tau_2,x_0)=p$. 
  Combining this
 with~\eqref{eq:ytre2} implies that
   we may actually assume that~$p-q \in \partial P^k_-$, so
\[
					p-q \not \in  P^k_+  ,
\]
and arguing just as above yields a  contradiction  that proves~\eqref{eq:pgt4}. 
%%%%%
\end{proof}

%%%%%%%%%%%%%%%%%
\begin{Lemma}\label{lemma:periodic}
%%%%%%%%%%%%%%%%%%%%
 Suppose that
	the   system~$\dot x=f(x)$ is strongly~$2$-cooperative.
%%%%%%
Let~$\gamma$ be a closed orbit corresponding to a periodic solution~$x(t+T,x_0)=x(t,x_0)$ 
for all~$t\geq 0$,
where~$T>0$ is the minimal period.
%%%%% 
Then  there exists an odd integer~$\ell \geq 1$ such that
\begin{align}
\label{eq:whatwe}
\ell-1 &\leq 
s^- ( x(t_2,x_0)-x(t_1,x_0) )\nonumber \\&
\leq s^+ ( x(t_2,x_0)-x(t_1,x_0) )  \leq \ell  
\end{align}
%%%%
for all   
$0< t_2-t_1 <T$.
%%%%%%%%%%%%%%%%%%%%%%%%%%%%% 
\end{Lemma}

\begin{proof} 
%%%
 We consider several cases. 

	\noindent \emph{Case 1.} 
	%%%%%
	Suppose that there exist~$\tau \geq 0$ and~$k\in\{1,2\}$ such that~$\dot x(\tau,x_0) \in P^k_- $. 
	Then~\eqref{eq:psdv} implies that~$\dot x(\tau,x_0) \in P^2_- $  (i.e.~$x(t,x_0)$ is pseudo-ordered). 
Lemma~\ref{lemma:periodicsol2} implies that
any two distinct points~$p,q \in \gamma$ satisfy
$p-q  \in P^2_+$,   
so~\eqref{eq:whatwe} holds with~$\ell=1$. 

		%%%%%%%%%%%%%%%%%%%%
	\noindent \emph{Case 2.}
		%%%%%%%%%%%%%%%%%%%%
		Suppose that Case~1 does not hold, and    that there exist~$\tau \geq 0$ and~$k\in\{3,4\}$ such that~$\dot x(\tau,x_0) \in P^k_- $.
	Then  Lemma~\ref{lemma:periodicsol2} implies that
	for any~$p,q \in \gamma$ with~$p\not =q$ we have
\be\label{eq:s09}
   p-q \in P^4_+ . 
\ee
%%%%%%%
	Since we assume that Case~1 does not hold,~$s^-(\dot x(t,x_0)) >1$ for all~$t$, so 
	Lemma~\ref{lemma:periodicsol2} implies that
	\[
	 p-q\not \in P^2_-.
	\]
	Combining this with~\eqref{eq:s09}, we  conclude that
$2\leq s^-(p-q) \leq s^+(p-q)\leq 3$,   
%%%%
so~\eqref{eq:whatwe} holds with~$\ell=3$. 
		%%%%%
		
The next case is when 	 Cases~1 and~2 do not hold, and   
	there exist~$\tau \geq 0$ and~$k\in\{5,6\}$ such that~$\dot x(\tau,x_0) \in P^k_- $. A similar argument in this case (and all other cases) 
  completes   the proof. 
\end{proof}

The next result describes an important application 
 of Lemma~\ref{lemma:periodic}.
We use~$e^i\in\R^n$ to denote the~$i$th canonical vector
in~$\R^n$.
 Let~$W^{1n}:= \text{span}\{ e^1, e^n  \}$. Clearly, this is a
    two-dimensional subspace that is contained in~$  P^2_-$.
 %%%%%%%%%%%%%%% 
\begin{Lemma}\label{lem:progtr}
%%%
Suppose that the conditions
 in Lemma~\ref{lemma:periodic} hold.   
 Then 
the orthogonal projection of~$\gamma$ to~$W^{1n}$ is one-to-one.  
%%%%%%%%%%%%%%%%%
\end{Lemma}

\begin{proof}
%%%
Seeking a contradiction, assume that there
 exist~$ p,q\in \gamma$, with~$p\not =q$,  such that
\[
p_1-q_1=p_n-q_n=0 .
\]					
It is easy to see that this implies that~$s^+(p-q)\geq 2+s^-(p-q) $. However, this contradicts~\eqref{eq:whatwe}.
\end{proof}

	We can now describe the proof of main result. 
	\begin{proof}[Proof of Thm.~\ref{thm:2dim}]
	%%%%
	  Using the fact that strongly $2$-cooperativity  implies 
		strongly  $2i$-cooperativity for every~$i$, we showed that
		any periodic solution (and not only pseudo-ordered periodic solutions)
		can be projected to a two-dimensional subspace in a one-to-one way. 
		Now the  remainder of  the   proof of Thm.~\ref{thm:2dim}
		follows from   the proof of Thm.~\ref{thm:sancg}, 
		\updt{which appears in~\citep{sanchez2009cones} as Thm. 1.}
	 
	%%%%
	\end{proof}
	
%%%%%%%%%%%%%%%%%%%%%%%%%%%%%%%%%%%%%%%%%%%%%%%%%%%%%%%%%%%% 
%%%%%%%%%%%%%%%%%%%%%%%%%%%%%%%%
%%%%%%%%%%%%%%%%%%%%%%%%%%%%%%%%%%%%%%
\section{Conclusion }\label{sec:conc}
%%%%%%%%%%%%%%%%%%%%%%%%%%%%%%%%%%%%%%%%%%

Positive  dynamical  systems are typically 
defined as systems  whose flow 
 maps~$\R^n_+$ to~$\R^n_+$. In fact,  the flow  maps 
the~$1$-solid cone~$P^1_-=\R^n_+ \cup \R^n_-$ to itself.  The important asymptotic properties of positive systems follow from 
the fact that they admit  an invariant~$1$-solid cone. Roughly speaking, this implies that
 a  trajectory can be projected to  a one-dimensional 
subspace and that this projection is generically one-to-one. 
Hence almost every    trajectory that remains in a compact set     converges to an
equilibrium. 

%%%%%
The reason that~$\R^n_+$ (and~$\R^n_-$) are also invariant sets of positive systems
is only because the
only way to cross from~$\R^n_+$ to~$\R^n_-$ (or vice versa) is through
the origin. 
%%%%%

Using tools  from the theory of TP matrices and totally positive
 differential systems, we  
introduced a  
 generalization  
called a~$k$-positive   LTV. This is   a  system in the form~$\dot x(t)=A(t) x(t)$
 whose dynamics
 maps the~$k$-solid cone~$P^k_-$
to itself. 
We showed how this property can be analyzed using 
 the  minors of order~$k$ of the transition matrix of the~LTV.  
In  the case where the matrix in the~LTV is a continuous function of time
we derived 
a necessary and sufficient condition for~$k$-positivity  
in terms of the~$k$'th additive compound of the matrix~$A(t)$.
This condition is straightforward  to verify and, in particular, does not require
to calculate the corresponding transition matrix. 
We also 
 provided an  explicit description of every set~$P^k_-$ as the union  of certain convex cones. 

The results   for~LTVs were applied
to define and analyze  $k$-cooperative   
nonlinear time-varying dynamical systems, that is,
systems with a $k$-positive variational system. Our results provide new tools for the analysis of nonlinear dynamical systems.

	We  believe  that out results can be extended in several interesting directions. 
	First,
\updt{the} theory of positive and cooperative systems has been applied to 
many types of  
dynamical systems including those described by ODEs, PDEs, systems with time-delay, 
difference equations, and more.
 A promising direction for further research is to extend
the notion and applications 
of~$k$-positivity and~$k$-cooperativity to 
additional types of dynamical systems, such as those mentioned above,
and to dynamical systems that evolve on manifolds~\citep{MOSTAJERAN20177439}. 
%%%%%%%%%
Another possible research direction is 
the extension of~$k$-positivity to \emph{control systems}.
   
We analyzed here $k$-positivity with respect to the set~$P^k_-$. 
Obviously, it is possible that~$\dot x=Ax$ is not $k$-positive
yet there exists an invertible matrix~$T $ such that the dynamical system for~$y(t):=Tx(t)$ is $k$-positive. A systematic analysis of when this is possible 
can greatly extend the applications of the theory. 

Finally,  
Example~\ref{exa:sminuin_2} illustrates that although
we can write~$P^k_-$ as a union of the
convex sets~$C^k_-(v^i)$ and~$-C^k_-(v^i)$, 
we do not know
how the solution actually evolves from one convex set to another. A
deeper understanding of the  
sign changes  that can  take place along the solution 
may yield stronger analysis results.

 \section*{Acknowledgments}
%%%%
We thank Rami Katz and Eduardo D. Sontag 
for many helpful comments. We are grateful to the anonymous reviewers  and the Associate Editor for a very helpful feedback.  

\section*{Appendix} 
%%%%%%%%%%%
{\sl Proof of Thm.~\ref{thm:srkew}.}
%%%%%%%%%%%%%%%%%%%%%%%%%%%%%%%%%%%%%%%%%%%%%
Suppose that~$A$ is nonsingular and~$SR_k$. 
For~$y \in\R$, let~$F(y)$  denote the~$n\times n$ matrix
whose~$i,j$ entry is~$\exp(- (i-j)^2 y) $. For example, for~$n=3$, 
\[
F(y)=\begin{bmatrix} 
                             1& \exp(-   y)& \exp(- 4  y)\\
														  \exp(-   y)&1& \exp(-    y)\\
														 \exp(-4   y)& \exp(-    y)&1
\end{bmatrix} .
\]
It is well-known that~$F(y)$ is~TP for all~$y>0$~\cite[Ch.~II]{gk_book},
and clearly~$\lim_{y\to\infty}F(y)=I$.
Fix~$y>0$ and let~$F:=F(y)$, and~$B:=F A$.
Let~$\alpha,\beta$
denote two sets of~$k$ integers~$1\leq i_1<\dots<i_k\leq n$ and~$1\leq j_1<\dots<j_k\leq n$, respectively.
The Cauchy-Binet formula yields
\[
B(\alpha|\beta) =\sum_{\gamma} F (\alpha|\gamma)  A(\gamma | \beta),
\]
where the sum is over all~$\gamma=\{p_1,\dots,p_k\}$, with~$ 1\leq p_1<\dots< p_k\leq n $.
Using the facts that~$F$ is TP, the  minors of order~$k$ of~$A$
are either all nonnegative or all nonpositive and they are not all zero (as~$A$ is nonsingular), we conclude
that~$B$ is~$SSR_k$.
Now pick~$x\in\R^n$ such that~$s^-(x)\leq k-1$. If~$x=0$ then clearly~$s^-(Bx) \leq   k-1$. If~$x\not =0$ then Thm.~\ref{thm:gtre}
implies that~$  s^+(Bx)\leq k-1$. 
We conclude that~$s^-(Bx) \leq   k-1$. Taking~$y \to \infty$ and using the fact that~$P^k_-$ is closed
yields~\eqref{eq:dcr}. 

To prove the converse implication, suppose that condition~\eqref{cond:onedip}  holds, that is,
for any~$x\in\R^n $ with~$ s^-(x) \leq k-1$, we have 
$
%%%%%
					s^-(Ax)\leq k-1$.
%%%%%%
Pick~$x\in \R^n\setminus\{0\}$ with~$s^-(x)\leq k-1$. Since~$A$ is nonsingular,~$Ax\not =0$. 
For any~$y>0$ the matrix~$F(y)$ is~TP,  so
\[
					s^+(F(y)Ax ) \leq s^-(Ax) ,
\]
and applying condition~\eqref{cond:onedip} yields
\[
					s^+(F(y)Ax ) \leq k-1.
%%%
\]
Thm.~\ref{thm:gtre} implies that~$F(y)A$ is~$SSR_k$. Taking~$y\to\infty$
and using continuity of the 
determinant,
we conclude that~$A$ is~$SR_k$. 
This completes the proof of Thm.~\ref{thm:srkew}.~\hfill{$\square$}

\bibliographystyle{abbrvnat}
\bibliography{MED2019_EYAL_bib}

\begin{thebibliography}{37}
\providecommand{\natexlab}[1]{#1}
\providecommand{\url}[1]{\texttt{#1}}
\expandafter\ifx\csname urlstyle\endcsname\relax
  \providecommand{\doi}[1]{doi: #1}\else
  \providecommand{\doi}{doi: \begingroup \urlstyle{rm}\Url}\fi

\bibitem[Alseidi et~al.(2019)Alseidi, Margaliot, and Garloff]{rola_spect}
R.~Alseidi, M.~Margaliot, and J.~Garloff.
\newblock On the spectral properties of nonsingular matrices that are strictly
  sign-regular for some order with applications to totally positive
  discrete-time systems.
\newblock \emph{J. Math. Anal. Appl.}, 474:\penalty0 524--543, 2019.

\bibitem[Arnaud(1998)]{Arnaud1998}
M.-C. Arnaud.
\newblock Le closing lemma en topologie $c^1$.
\newblock \emph{Memoires de la Societe Mathematique de France}, 74:\penalty0
  1--120, 1998.
\newblock URL \url{http://eudml.org/doc/94925}.

\bibitem[Ben-Avraham et~al.(2019)Ben-Avraham, Sharon, Zarai, and
  Margaliot]{CTPDS}
T.~Ben-Avraham, G.~Sharon, Y.~Zarai, and M.~Margaliot.
\newblock Dynamical systems with a cyclic sign variation diminishing property.
\newblock \emph{IEEE Trans.\ Automat.\ Control}, 65\penalty0 (3):\penalty0
  941--954, 2019.

\bibitem[Blanchini(1999)]{blanchini1999set}
F.~Blanchini.
\newblock Set invariance in control.
\newblock \emph{Automatica}, 35\penalty0 (11):\penalty0 1747--1767, 1999.

\bibitem[Byrnes(1999)]{Byrnes_global}
C.~I. Byrnes.
\newblock On the global analysis of linear systems.
\newblock In J.~Baillieul and J.~C. Willems, editors, \emph{Mathematical
  Control Theory}, pages 99--139, New-York, 1999. Springer-Verlag.

\bibitem[Elkhader(1992)]{Elkhader1992}
A.~S. Elkhader.
\newblock A result on a feedback system of ordinary differential equations.
\newblock \emph{J. Dyn. Diff. Equat.}, 4\penalty0 (3):\penalty0 399--418, 1992.

\bibitem[Fallat et~al.(2017)Fallat, Johnson, and Sokal]{fallat2017total}
S.~Fallat, C.~R. Johnson, and A.~D. Sokal.
\newblock Total positivity of sums, {H}adamard products and {H}adamard powers:
  Results and counterexamples.
\newblock \emph{Linear Algebra Appl.}, 520:\penalty0 242--259, 2017.

\bibitem[Fallat and Johnson(2011)]{total_book}
S.~M. Fallat and C.~R. Johnson.
\newblock \emph{Totally Nonnegative Matrices}.
\newblock Princeton University Press, Princeton, NJ, 2011.

\bibitem[Farina and Rinaldi(2000)]{farina2000}
L.~Farina and S.~Rinaldi.
\newblock \emph{Positive Linear Systems: Theory and Applications}.
\newblock John Wiley, 2000.

\bibitem[Feng et~al.(2017)Feng, Wang, and Wu]{feng2017}
L.~Feng, Y.~Wang, and J.~Wu.
\newblock Semiflows ``monotone with respect to high-rank cones'' on a {Banach}
  space.
\newblock \emph{SIAM J. Math. Anal.}, 49\penalty0 (1):\penalty0 142--161, 2017.

\bibitem[Fiedler(2008)]{fiedler_book}
M.~Fiedler.
\newblock \emph{Special Matrices and Their Applications in Numerical
  Mathematics}.
\newblock Dover Publications, Mineola, NY, 2 edition, 2008.

\bibitem[Gantmacher and Krein(2002)]{gk_book}
F.~R. Gantmacher and M.~G. Krein.
\newblock \emph{Oscillation Matrices and Kernels and Small Vibrations of
  Mechanical Systems}.
\newblock American Mathematical Society, Providence, RI, 2002.
\newblock Translation based on the~1941 {Russian} original.

\bibitem[Haag(2017)]{gunter2017}
G.~Haag.
\newblock \emph{Modelling with the Master Equation}.
\newblock Springer, Cham, Switzerland, 2017.

\bibitem[Horv{\'a}th et~al.(2016)Horv{\'a}th, Song, and
  Terlaky]{horvath2016invariance}
Z.~Horv{\'a}th, Y.~Song, and T.~Terlaky.
\newblock Invariance conditions for nonlinear dynamical systems.
\newblock In B.~Goldengorin, editor, \emph{Optimization and Its Applications in
  Control and Data Sciences}, volume 115 of \emph{Springer Optimization and Its
  Applications}, chapter~8, pages 265--280. Springer, 2016.

\bibitem[Johnson and Pena(2007)]{sign_pres_mats}
C.~R. Johnson and J.~M. Pena.
\newblock Matrices that preserve vectors of fixed sign variation.
\newblock \emph{Linear and Multilinear Algebra}, 55\penalty0 (6):\penalty0
  521--533, 2007.

\bibitem[Katz et~al.(2020)Katz, Margaliot, and Fridman]{rami_osci}
R.~Katz, M.~Margaliot, and E.~Fridman.
\newblock Entrainment to subharmonic trajectories in oscillatory discrete-time
  systems.
\newblock \emph{Automatica}, 116\penalty0 (108919), 2020.

\bibitem[Krasnoselskii et~al.(1989)Krasnoselskii, Lifshitz, and
  Sobolev]{pls_sobolev}
M.~A. Krasnoselskii, E.~A. Lifshitz, and A.~V. Sobolev.
\newblock \emph{Positive Linear Systems: The Method of Positive Operators}.
\newblock Heldermann Verlag, Berlin, 1989.

\bibitem[Kushel(2012)]{kushel2012cone}
O.~Y. Kushel.
\newblock Cone-theoretic generalization of total positivity.
\newblock \emph{Linear Algebra Appl.}, 436\penalty0 (3):\penalty0 537--560,
  2012.

\bibitem[Mallet-Paret and Smith(1990)]{poin_cyclic}
J.~Mallet-Paret and H.~L. Smith.
\newblock The {P}oincar{\'e}-{B}endixson theorem for monotone cyclic feedback
  systems.
\newblock \emph{J. Dyn. Differ. Equ.}, 2\penalty0 (4):\penalty0 367--421, 1990.

\bibitem[Margaliot and Sontag(2019{\natexlab{a}})]{Margaliot868000}
M.~Margaliot and E.~D. Sontag.
\newblock Compact attractors of an antithetic integral feedback system have a
  simple structure.
\newblock \emph{bioRxiv}, 2019{\natexlab{a}}.
\newblock \doi{10.1101/868000}.
\newblock URL \url{https://www.biorxiv.org/content/early/2019/12/08/868000}.

\bibitem[Margaliot and Sontag(2019{\natexlab{b}})]{margaliot2019revisiting}
M.~Margaliot and E.~D. Sontag.
\newblock Revisiting totally positive differential systems: A tutorial and new
  results.
\newblock \emph{Automatica}, 101:\penalty0 1--14, 2019{\natexlab{b}}.

\bibitem[Mostajeran and Sepulchre(2017)]{MOSTAJERAN20177439}
C.~Mostajeran and R.~Sepulchre.
\newblock Differential positivity with respect to cones of rank $k \geq 2$.
\newblock In \emph{Proc. 20th {IFAC} World Congress}, volume~50, pages
  7439--7444, Toulouse, France, 2017.

\bibitem[Muldowney(1990)]{muldo1990}
J.~S. Muldowney.
\newblock Compound matrices and ordinary differential equations.
\newblock \emph{The Rocky Mountain J. Math.}, 20\penalty0 (4):\penalty0
  857--872, 1990.

\bibitem[Oliva et~al.(1993)Oliva, Kuhl, and
  Magalh{\~a}es]{oliva1993diffeomorphisms}
W.~M. Oliva, N.~M. Kuhl, and L.~T. Magalh{\~a}es.
\newblock Diffeomorphisms of {$\R^n$} with oscillatory {Jacobians}.
\newblock \emph{Publicacions Matem{\`a}tiques}, pages 255--269, 1993.

\bibitem[Pinkus(1996)]{Pinkus1996}
A.~Pinkus.
\newblock Spectral properties of totally positive kernels and matrices.
\newblock In M.~Gasca and C.~A. Micchelli, editors, \emph{Total Positivity and
  its Applications}, pages 477--511. Springer Netherlands, Dordrecht, 1996.

\bibitem[Pinkus(2010)]{pinkus}
A.~Pinkus.
\newblock \emph{Totally Positive Matrices}.
\newblock Cambridge University Press, Cambridge, UK, 2010.

\bibitem[Rantzer and Valcher(2018)]{posi-tutorial}
A.~Rantzer and M.~E. Valcher.
\newblock A tutorial on positive systems and large scale control.
\newblock In \emph{{Proc.\ 57th IEEE Conf. on Decision and Control}}, pages
  3686--3697, Miami Beach, FL, USA, 2018.

\bibitem[Sanchez(2009)]{sanchez2009cones}
L.~A. Sanchez.
\newblock Cones of rank 2 and the {P}oincar{\'e}-{B}endixson property for a new
  class of monotone systems.
\newblock \emph{J. Diff. Eqns.}, 246\penalty0 (5):\penalty0 1978--1990, 2009.

\bibitem[Sandberg(1978)]{sandberg78}
I.~W. Sandberg.
\newblock On the mathematical foundations of compartmental analysis in biology,
  medicine, and ecology.
\newblock \emph{IEEE Trans. Circuits and Systems}, 25\penalty0 (5):\penalty0
  273--279, 1978.

\bibitem[Schwarz(1970)]{schwarz1970}
B.~Schwarz.
\newblock Totally positive differential systems.
\newblock \emph{Pacific J. Math.}, 32\penalty0 (1):\penalty0 203--229, 1970.

\bibitem[Smillie(1984)]{smillie}
J.~Smillie.
\newblock Competitive and cooperative tridiagonal systems of differential
  equations.
\newblock \emph{SIAM J. Math. Anal.}, 15:\penalty0 530--534, 1984.

\bibitem[Smith(1991)]{periodic_tridi_smith}
H.~L. Smith.
\newblock Periodic tridiagonal competitive and cooperative systems of
  differential equations.
\newblock \emph{SIAM J. Math. Anal.}, 22\penalty0 (4):\penalty0 1102--1109,
  1991.

\bibitem[Smith(1995)]{hlsmith}
H.~L. Smith.
\newblock \emph{Monotone Dynamical Systems: An Introduction to the Theory of
  Competitive and Cooperative Systems}, volume~41 of \emph{Mathematical Surveys
  and Monographs}.
\newblock Amer. Math. Soc., Providence, RI, 1995.

\bibitem[Song(2015)]{song2015optimization}
Y.~Song.
\newblock \emph{Optimization Theory and Dynamical Systems: Invariant Sets and
  Invariance Preserving Discretization Methods}.
\newblock PhD thesis, Lehigh University, 2015.

\bibitem[Sontag(1998)]{sontag_book}
E.~D. Sontag.
\newblock \emph{Mathematical Control Theory: Deterministic Finite Dimensional
  Systems}.
\newblock Springer, New York, 2 edition, 1998.

\bibitem[Walter(1997)]{Walter1997}
W.~Walter.
\newblock On strongly monotone flows.
\newblock \emph{Annales Polonici Mathematici}, 66\penalty0 (1):\penalty0
  269--274, 1997.

\bibitem[Weiss and Margaliot(2018)]{Eyal_Smiliie}
E.~Weiss and M.~Margaliot.
\newblock A generalization of {S}millie's theorem on strongly cooperative
  tridiagonal systems.
\newblock In \emph{{Proc.\ 57th IEEE Conf. on Decision and Control}}, pages
  3080--3085, Miami, FL, 2018.

\end{thebibliography}

	%%%\bibliographystyle{abbrvnat}
%%\bibliography{tpdsbib}

%%%%%%%%%%%%%%%%%
\begin{comment}
%%%%%%%%%%%%%%%%%%%
MATLAB CODE FOR EXAMPLES
%%%%%%%%%%%%%%%%%%%%%%%%%

hold off
A=[-1 2 -2 1; 3 0 1 -1; ............
     -4 1.5  2 4; 1 -1 2  5];
A
 x0=[   -0.27   -0.85   -0.45    0.09]';
 x0=x0+1*[ 0.61    0.31   -0.61     0.4]';
 
for t=0:.01:2.5
xt=expm(A*t)*x0;
     num=0;
if xt(1)*xt(2)*xt(3)*xt(4) ~= 0
     % no zero entries
     if xt(1)*xt(2)<0 num=num+1; end
     if xt(2)*xt(3)<0 num=num+1; end
     if xt(3)*xt(4)<0 num=num+1; end
end
plot(t,num,'.k'); hold on;
end
grid on;
xlabel('$t$','interpreter','latex', .....
     'fontsize', 16);
aa=axis;
axis([ aa(1) aa(2) 0  aa(4)]);

SECOND EXAMPLE
-----------------

\end{comment}

%\end{multicols} %just before \end{document}
 \end{document}